\documentclass[aps,prd,widetext,showpacs,preprintnumbers,superscriptaddress]{revtex4}

\usepackage{graphicx}
\usepackage{amssymb}
\usepackage{amsmath}
\usepackage{dcolumn}
\usepackage{bm}
\usepackage{natbib}
\usepackage{multirow}
\usepackage{aas_macros}

\begin{document}

\newcommand{\rem}[1]{{\bf #1}}

\preprint{ICRR-Report-559-2009-21,\ IPMU10-0024}

\title{Gravitational Waves from Collapsing Domain Walls}

\author{Takashi Hiramatsu}%
\affiliation{Institute for Cosmic Ray Research, The University of Tokyo, 5-1-5 Kashiwa-no-ha, Kashiwa City, Chiba 277-8582, Japan}
\author{Masahiro Kawasaki}
\affiliation{Institute for Cosmic Ray Research, The University of Tokyo, 5-1-5 Kashiwa-no-ha, Kashiwa City, Chiba 277-8582, Japan} 
\affiliation{Institute for Physics and Mathematics of the Universe, The University of Tokyo, 5-1-5 Kashiwa-no-ha, Kashiwa City, Chiba 277-8582, Japan}
 \author{Ken'ichi Saikawa}
 \affiliation{Institute for Cosmic Ray Research, The University of Tokyo, 5-1-5 Kashiwa-no-ha, Kashiwa City, Chiba 277-8582, Japan}


\begin{abstract}
We study the production of gravitational waves from cosmic domain walls 
created during phase transition in the early universe. We investigate the 
process of formation and evolution of domain walls by running three 
dimensional lattice simulations. If we introduce an approximate discrete 
symmetry, walls become metastable and finally disappear. 
This process might occur 
by a pressure difference between two vacua if a quantum tunneling is neglected.
We calculate the 
spectrum of gravitational waves produced by collapsing metastable domain 
walls. Extrapolating the numerical results, we find that the signal of 
gravitational waves produced by domain walls whose energy scale is around 
$10^{10}$-$10^{12}$GeV will be observable in the next generation 
gravitational wave interferometers.
\end{abstract}

\maketitle

\section{\label{sec1}Introduction}
The spontaneous symmetry breaking is one of the most important ingredients of 
the modern particle physics. Not only it describes the unification of forces 
acting on the elementary particles, but also it has rich consequences in the 
early universe. One is the occurrence of primordial phase 
transitions followed by the formation of topological defects \cite
{1976JPhA....9.1387K}. Especially, if there was the spontaneous breaking of a 
discrete symmetry, the two dimensional surface-like defects, called domain 
walls, would be formed in the early era of the universe (for a review of 
domain walls and other topological defects, see \cite{1994csot.book.....V}). 
Although discrete symmetries naturally arise in many particle physics models, 
the existence of stable domain walls is disfavored by cosmological 
considerations. Indeed, if domain walls survived until the present time, the 
energy density of walls would eventually come to dominate the total energy 
density of the universe, causing the fast expansion of the universe to affect 
the galaxy formation or primordial nucleosynthesis \cite{1989PhRvD..39.1558G}. 
Moreover, the presence of domain walls would cause the excessive anisotropy in 
the cosmic microwave background observed today, which rules out the existence 
of stable domain walls with the symmetry breaking scale $\eta\gtrsim$1MeV 
\cite{1974JETP...40....1Z}.

However, it is still possible to consider the existence of {\it unstable} 
domain walls which decay early enough not to lead cosmological disasters. One 
way to introduce the instability of walls is to make discrete symmetry only 
approximate and tilt the potential so as to lift the degeneracy of vacua. This 
might be raised by nonrenormalizable operators suppressed by the cutoff scale 
($\sim$  Planck scale)~\cite{1994PhRvD..49.2729R}. The idea of an approximate 
discrete symmetry as a mechanism to remove the domain wall problem has been 
studied by several authors~\cite{1989PhRvD..39.1558G,1981PhRvD..23..852V,
1982PhRvL..48.1156S,1984PhLB..143..351M}. Also, there is another possibility 
to create unstable domain walls due to a postinflationary nonthermal phase 
transition~\cite{1995PhRvD..51.5456L,1996PhRvD..53.4237C}.

In this paper we investigate the production of gravitational waves from the 
annihilating process of unstable domain walls. 
There are early works studying the bubble collisions in the early universe~\cite{1982PhRvD..26.2681H} 
and the possibility of generating gravitational waves~\cite{1990PhRvL..65.3080T,1992PhRvD..45.4514K,1992PhRvL..69.2026K,1993PhRvD..47.4372K,1994PhRvD..49.2837K}.
In these works, gravitational waves are considered to be generated by the collision of spherically 
expanding bubbles, which are created during first order phase transitions (i.e. the quantum 
tunneling from the false vacuum to the true vacuum). Instead, in this paper 
we assume that the transition from one vacuum to another proceeds via a
realignment of the classical field rather than a quantum tunneling.
The production of gravitational waves from such a process was originally 
calculated in~\cite{1998PhRvL..81.5497G}. However, 
in~\cite{1998PhRvL..81.5497G} the radiated energy is calculated only in the 
simple configuration such as situations with axial symmetry, and only the 
total intensity of gravitational waves was estimated. Instead, our purpose is 
to study the production of gravitational waves in more realistic situations 
(i.e. to follow from phase transition and to consider generic configuration of 
the scalar field), and calculate the spectrum of gravitational waves.

The gravitational wave is the powerful tool for studying cosmology and high 
energy physics (for reviews, 
see~\cite{2000PhR...331..283M,Maggiore2008,2009LRR....12....2S}). 
There are various ongoing and planned experiments on the gravitational 
waves. As for the ground-based experiment, 
LIGO~\cite{1992Sci...256..325A} is running, and LCGT~\cite{2002CQGra..19.1237K} is planned. And space-borne interferometers 
such as LISA~\cite{LISA}, BBO \cite{2005PhRvD..72h3005C}, and 
DECIGO~\cite{2006CQGra..23S.125K} are planned to be launched in the future. 
Furthermore, it has been started to plan the ground-based interferometer 
with improved sensitivity, ``Einstein Telescope(ET)''~\cite{ET} 
in Europe recently. 
In this paper we discuss whether the stochastic gravitational wave backgrounds 
produced by domain walls are detectable in these experiments. If it is 
detectable and the predicted form of the spectrum is distinctive, we expect that
the gravitational wave spectrum from domain walls can be another probe for 
high energy physics. One example is in the theory with supersymmetry. One may 
be able to measure the parameter of the theory such as the mass of the 
gravitino, fermionic partner of the graviton, by gravitational waves from 
annihilation of $Z_{2N}$ domain walls \cite{2008PhLB..664..194T}.

The dynamics of domain walls is not easy to investigate because of the
nonlinear nature of the evolution equations.  Thus it is necessary to
use numerical calculations to follow the evolution of domain walls with
arbitrary structures. The first complete numerical study was performed
by Press, Ryden, and Spergel~\cite{1989ApJ...347..590P} by using the
modified field equation for the scalar field (called the PRS
algorithm). Also, Kawano~\cite{1990PhRvD..41.1013K} simulated the
evolution of domain walls by using the thin wall effective equation of
motion (similar to the Nambu-Goto
equation~\cite{1970sqm..conf..269N,1971PThPh..46.1560G} for cosmic
strings).  Later, several authors investigated further by up-dated high
resolution
simulations~\cite{2003PhRvD..68j3506G,2005PhRvD..71h3509O,2005PhLB..610....1A}
and the numerical simulations for unstable domain walls are performed
in~\cite{1996PhRvD..53.4237C,1997PhRvD..55.5129L}. However, although
these numerical simulations successfully demonstrate the macroscopic
evolution of domain wall networks, they put an unphysical assumption
about the width of a domain wall, and we cannot use their computational
method since this method may invalidates the estimation of the energy of
gravitational waves (see Section~\ref{sec4A}). Therefore, we solve the
field equations directly rather than rely on the previous
methodologies. This gives us severe technical restrictions in choosing
parameters of the simulations.

This paper is organized as follows. In Section~\ref{sec2}, we shortly review 
the dynamics of domain wall networks. In Section~\ref{sec3}, we describe the 
formalism to calculate classical gravitational wave spectra radiated by 
the scalar field in the systematic way. Then, after a short discussion about 
the difficulty in numerical simulations, we present the results of our 
numerical simulations in Section~\ref{sec4}. We combine the results of 
numerical calculations with analytic estimations, and discuss the 
observability of the spectrum by extrapolating them in Section~\ref{sec5}. 
Finally we conclude in Section~\ref{sec6}.

\section{\label{sec2}Domain Wall Dynamics}

\subsection{\label{sec2A}The Model of Domain Walls}

We consider the model of the real scalar field with the potential
\begin{equation}
   V(\phi) = \frac{\lambda}{4}(\phi^2-\eta^2)^2 
   + \epsilon\eta\phi\left(\frac{1}{3}\phi^2-\eta^2\right) 
   + \frac{\lambda}{8}T^2\phi^2. \label{eq2-1}
\end{equation}
The first term is the usual double well potential. In the absence of 
the second term, this model has a discrete $Z_2$ symmetry corresponding to 
the transformation $\phi\to-\phi$. The second term is introduced in order to 
lift the degeneracy of the two vacua and make the discrete symmetry 
approximate. In this model, the scalar field is assumed to be thermal 
equilibrium at the first stage of their evolution, and we introduce the 
finite temperature correction in the last term of the potential. When 
finite temperature effects become negligible, the potential has two 
minima $\phi=\pm\eta$ and the formation of domain walls occurs. 
The difference of the energy density between two minima is 
$\Lambda=4\epsilon\eta^4/3$ and the height of the potential barrier 
is $\Delta V \simeq \lambda\eta^4/4$. The dimensionless constant 
parameter $\epsilon$ is called ``bias'', which makes domain walls unstable. 

In the numerical simulations which we will describe later, we solve the classical 
filed equation with the potential given by Eq.~(\ref{eq2-1}). This corresponds to 
the fact that we consider only a quiescent second order transition.
If there are more violent first order transitions via a quantum tunneling, 
the gravitational wave signature might be enhanced significantly.
Such a quantum treatment is, however, beyond the scope of this paper.

In the following two subsections, we briefly review the property of 
the evolution of domain wall networks.

\subsection{\label{sec2B}Scaling Solutions}

The well known property of the evolution of domain wall networks is that 
there exists the phase of ``scaling'', in which the typical length scales 
such as the wall curvature radius $R$ and the distance of two neighboring 
walls $L$ are given by Hubble radius,
\begin{equation}
   R \sim L \sim H^{-1} \sim t. 
   \label{eq2-2}
\end{equation}
This property is originally studied in~\cite{1989ApJ...347..590P} 
numerically, and by other groups in~\cite{1996PhRvD..53.4237C,
1997PhRvD..55.5129L,2003PhRvD..68j3506G,2005PhRvD..71h3509O,
2005PhLB..610....1A}. 
However, the analytic modeling of the scaling solution is found 
in~\cite{1996PhRvL..77.4495H,2003PhRvD..68d3510H,2005PhRvD..72h3506A}. 

When domain wall networks are in scaling regime, the energy density of 
domain walls is given by
\begin{equation}
   \rho_w \sim \sigma R^2/R^3 \sim \sigma/t, 
   \label{eq2-3}
\end{equation}
where $\sigma = 2\sqrt{2\lambda}\eta^3/3$ is the tension of the wall. 
Eq.~(\ref{eq2-3}) is also expressed as
\begin{equation}
   A/V \propto \tau^{-1}, 
   \label{eq2-4}
\end{equation}
where $A/V$ is the comoving area density of the wall, and $\tau$ is the 
conformal time: $d\tau = dt/a$. If we assume radiation dominated universe, 
$A/V$ is proportional to $t^{-1/2}$, which is a criterion to distinguish 
whether walls are in scaling regime. 

\subsection{\label{sec2C}The Role of the Bias}

There are two effects caused by the bias in the potential. One is the 
bias of two vacua chosen during the phase transition, and another is the 
volume pressure which shrinks the false vacuum domains and eliminates 
domain walls. The cosmological scenario with biased discrete symmetry 
is investigated in~\cite{1989PhRvD..39.1558G} and here we follow the 
analysis of it.

Let us denote the probability of having a scalar field fluctuation during 
the phase transition end up in plus vacuum as $p_+$ and in minus vacuum as 
$p_-$. If there exists the bias, the ratio of these two probabilities is 
given by
\begin{equation}
   \frac{p_+}{p_-} = \exp\left(-\frac{\Delta F}{T}\right). 
   \label{eq2-5}
\end{equation}
Here, $\Delta F = \Lambda\times\xi_{\mathrm{corr}}^3$ is the difference 
of the free energy between two vacua, and $\xi_{\mathrm{corr}}$ is the 
correlation length of the scalar fields. If we estimate $T$ as Ginzburg 
temperature~\cite{1976JPhA....9.1387K,1989PhRvD..39.1558G}, given by 
$T_G \simeq \Delta V \times \xi_{\mathrm{corr}}^3 \simeq 
\lambda\eta^4\xi_{\mathrm{corr}}^3/4$, the above equation can be written 
as
\begin{equation}
   \frac{p_+}{p_-} = \exp\left(-\frac{\Lambda}{\Delta V}\right) 
   \simeq \exp\left(-\frac{16}{3}\frac{\epsilon}{\lambda}\right). 
   \label{eq2-6}
\end{equation}
Therefore, there exists the bias between two vacua if $\epsilon \ne 0$.
The spatial distribution of two vacua after phase transition has been
studied by using percolation theory \cite{1979PhR....54....1S}.  It has
been shown that if the probability $p_+$ (or $p_-$) is bigger than a
critical value $p_c$, an infinite plus (minus) cluster appears in the
space. If we approximate the space as a three dimensional cubic lattice,
$p_c=0.311$ \cite{1979PhR....54....1S}, and by requiring both $p_+$ and
$p_-$ are bigger than $p_c$, we obtain
\begin{equation}
   \epsilon < 0.15\lambda. 
   \label{eq2-7}
\end{equation}
If the condition given by Eq. (\ref{eq2-7}) is satisfied, the infinite
size of domain walls would appear after phase transition.  

After the formation of domain walls, there are two forces acting on the walls.
The first force is a surface tension, which is given by $p_T \simeq
\sigma/R \simeq 2\sqrt{2\lambda}\eta^3/3R$ (as a pressure).  It smooths
out the small scale structure on the wall and straightens the wall up to
the horizon scale. The second force is a volume pressure, which is given
by $p_V \simeq \Lambda = 4\epsilon\eta^4/3$.  It accelerates the wall
against the false vacuum regions due to the energy difference between
the two vacua and collapses the wall.  The evolution of domain wall
networks depends on when and which of the two forces dominate.  The
volume pressure dominates when $p_V \simeq p_T$ is satisfied, and the
corresponding wall curvature radius is
\begin{equation}
   R \simeq \sqrt{\frac{\lambda}{2}}(\epsilon\eta)^{-1}. 
   \label{eq2-8}
\end{equation}
If we require that the wall has been straighten up to the horizon scale
before the volume pressure becomes effective, we obtain the condition
$R>H^{-1}(t_c)$ where $R$ is given by Eq. (\ref{eq2-8}) and $H(t_c)$ is
the Hubble parameter at the phase transition ($T=T_c=2\eta$). This
condition gives an upper bound for $\epsilon$:
\begin{equation}
   \epsilon < 4.7\times (\lambda g_*)^{1/2}\frac{\eta}{M_P}, 
    \label{eq2-9}
\end{equation}
where $g_*$ is the number of relativistic degrees of freedom at $T_c$
and $M_P$ is the Planck mass. On the other hand, a requirement that
domain walls do not dominate the energy density of the universe gives a
lower bound for $\epsilon$. When domain wall networks are in scaling
regime, the energy density of domain walls is estimated as $\rho_w
\simeq \sigma/H^{-1} \simeq \sqrt{2\lambda}\eta^3/3t $. If we assume the
radiation dominated universe, the total energy density of the universe
is given by $\rho \simeq 3M_P^2/32\pi t^2$. The domain walls start to
dominate the universe ($\rho_w/\rho \simeq 1$) when $t=t_{WD}$, where
\begin{equation}
   t_{WD} \simeq \frac{9M_P^2}{64\pi}\sqrt{\frac{2}{\lambda}}\eta^{-3}. 
    \label{eq2-10}
\end{equation}
Meanwhile, the typical time scale of the collapse of the wall network is
given by the condition $H^{-1}(t_{\mathrm{dec}}) =
2t_{\mathrm{dec}}\simeq R$ where $R$ is given by Eq. (\ref{eq2-8}):
\begin{equation}
   t_{\mathrm{dec}} \simeq 
    \frac{1}{2}\sqrt{\frac{\lambda}{2}}(\epsilon\eta)^{-1}. 
    \label{eq2-11}
\end{equation}
Therefore, the condition that domain walls collapse before wall
domination is $t_{\mathrm{dec}}<t_{WD}$, from which we obtain
\begin{equation}
   \epsilon > \frac{16\pi}{9}\lambda\left(\frac{\eta}{M_P}\right)^2.
    \label{eq2-12}
\end{equation}

To sum up, the domains with infinite size appear if the condition given
by Eq.~(\ref{eq2-7}) is satisfied, and the domain walls collapse before wall domination if $\epsilon$
satisfies Eqs.~(\ref{eq2-9}) and (\ref{eq2-12}). The decay time of
the domain walls is given by Eq. (\ref{eq2-11}). 

\section{\label{sec3}calculation of Gravitational Waves}

In this section, we describe how to calculate the spectrum of the stochastic
gravitational wave backgrounds produced by domain walls. We use the
formalism of ``Green's function method'' developed by Dufaux et
al.~\cite{2007PhRvD..76l3517D,2009JCAP...03..001D}. Originally this
formalism was applied to the production of gravitational waves from
preheating \cite{1994PhRvL..73.3195K,1997PhRvD..56.3258K} after
inflation. However, this framework is applicable to general scalar
fields in expanding universe, especially the Higgs field forming domain
walls considered here. In the following, we summarize the results of
\cite{2007PhRvD..76l3517D}.

Assuming a spatially flat Friedmann-Robertson-Walker background,
gravitational waves are represented by the spatial metric perturbation
\begin{equation}
    ds^2 = a^2(\tau)[-d\tau^2+(\delta_{ij}+h_{ij})dx^idx^j],
     \label{eq3-1}
\end{equation}
where $h_{ij}$ satisfies the condition $\partial_ih_{ij}=h^i_i=0$. The
evolution of $h_{ij}$ is described by the linearized Einstein equation
\begin{equation}
   \ddot{h}_{ij} + 3H\dot{h}_{ij} -\frac{\nabla^2}{a^2}h_{ij} 
    = 16\pi GT^{TT}_{ij}, 
    \label{eq3-2}
\end{equation}
where $T^{TT}_{ij}$ is the transverse-traceless part of the
stress-energy tensor, which is computed by applying the projection
operator in the momentum space
\begin{eqnarray}
   T^{TT}_{ij}(\tau,{\bf k}) &=& \Lambda_{ij,kl}(\hat{k})
    T_{ij}(\tau,{\bf k}) =  \Lambda_{ij,kl}(\hat{k}) 
    \{\partial_k\phi\partial_l\phi\}(\tau,{\bf k}), 
    \label{eq3-3}\\
    \Lambda_{ij,kl}(\hat{k}) &=& P_{ik}(\hat{k})P_{jl}(\hat{k}) 
     - \frac{1}{2}P_{ij}(\hat{k})P_{kl}(\hat{k}), 
     \label{eq3-4}\\
    P(\hat{k}) &=& \delta_{ij}-\hat{k}_i\hat{k}_j, 
     \label{eq3-5}
\end{eqnarray}
where $\hat{k}={\bf k}/|{\bf k}|$ and
$\{\partial_k\phi\partial_l\phi\}(\tau,{\bf k})$ is the Fourier
transform of $\partial_k\phi(\tau,{\bf x})\partial_l\phi(\tau,{\bf x})$.

Suppose that the source term is nonzero during the interval $\tau_i \le
\tau \le \tau_f$. In this case, the equation of motion (\ref{eq3-2}) is
solved by using the Green's function in $\tau_i \le \tau \le \tau_f$,
and the solution can be matched to the source-free solution in $\tau \ge
\tau_f$. In the radiation dominated universe, the result
is~\cite{2007PhRvD..76l3517D}
\begin{equation}
    \bar{h}_{ij}(\tau,{\bf k}) = A_{ij}({\bf k})\sin[k(\tau-\tau_f)] 
     + B_{ij}({\bf k})\cos[k(\tau-\tau_f)]\qquad 
     \mathrm{for}\qquad \tau \ge \tau_f, 
     \label{eq3-6}
\end{equation}
where
\begin{eqnarray}
    A_{ij}({\bf k}) &=& \frac{16\pi G}{k}
     \int^{\tau_f}_{\tau_i}d\tau'\cos[k(\tau_f-\tau')]
     a(\tau')T^{TT}_{ij}(\tau',{\bf k}), 
     \nonumber\\
    B_{ij}({\bf k}) &=& \frac{16\pi G}{k}
     \int^{\tau_f}_{\tau_i}d\tau'\sin[k(\tau_f-\tau')]
     a(\tau')T^{TT}_{ij}(\tau',{\bf k}), 
     \label{eq3-7}
\end{eqnarray}
and $\bar{h}_{ij}=a(\tau)h_{ij}$.

The energy density of gravitational waves is given by the spatial
average of the product of $\dot{h}_{ij}$ (see e.g.~\cite{Maggiore2008})
\begin{equation}
   \rho_{\mathrm{gw}} = \frac{1}{32\pi G}
    \langle \dot{h}_{ij}(t,{\bf x})\dot{h}_{ij}(t,{\bf x}) \rangle 
    = \frac{1}{32\pi Ga^4}\frac{1}{V}
    \int \frac{d^3k}{(2\pi)^3}\bar{h}_{ij}'(\tau,{\bf k})
    \bar{h}_{ij}'^*(\tau,{\bf k}), 
    \label{eq3-8}
\end{equation}
where a prime denotes a derivative with respect to the conformal time
$\tau$ and $V$ is the comoving volume. Strictly speaking, there are
other contributions for the right hand side of Eq. (\ref{eq3-8}): one is
cross terms between $\bar{h}_{ij}$ and $\bar{h}_{ij}'$ which turn out to
vanish under the reality condition of $\bar{h}_{ij}(\tau,{\bf x})$, and
another is the term proportional to $H^2$ which is ignorable if we
consider only the modes inside the horizon at present time $k/a_0 \gg
H_0$. Substituting Eq. (\ref{eq3-6}) into Eq. (\ref{eq3-8}) and taking
the time average over a period of the oscillations caused by the sine
and cosine terms in Eq. (\ref{eq3-6}) to eliminate these oscillations,
we find
\begin{equation}
   \rho_{\mathrm{gw}}(t) = \frac{4\pi G}{a^4(t)V}
    \int\frac{d^3k}{(2\pi)^3}\sum_{i,j}
    \left\{\left|\int^{\tau_f}_{\tau_i}d\tau'\cos(k\tau')a(\tau')
    T^{TT}_{ij}(\tau',{\bf k})\right|^2 
    + \left|\int^{\tau_f}_{\tau_i}d\tau'\sin(k\tau')a(\tau')
    T^{TT}_{ij}(\tau',{\bf k})\right|^2 \right\}. 
    \label{eq3-9}
\end{equation}

The present spectrum is represented by the fraction of energy density
against the total energy density of the universe per logarithmic
frequency interval
\begin{equation}
   \Omega_{\mathrm{gw}}h^2 = \frac{h^2}{\rho_{c,0}}
    \frac{d\rho_{\mathrm{gw}}(t_0,f)}{d\ln f} 
    = \frac{h^2}{\rho_{c,0}}\frac{d\rho_{\mathrm{gw}}(t_0,k)}{d\ln k},
    \label{eq3-10} 
\end{equation}
where $f$ is the frequency of gravitational waves, $\rho_{c,0}$ is the
critical energy density today, and $h$ is the renormalized Hubble
parameter: $H_0=100h$kmsec$^{-1}$Mpc$^{-1}$. Taking account of the fact
that energy density of gravitational waves redshifts as
$\rho_{\mathrm{gw}}\propto a^{-4}$, and from Eqs.~(\ref{eq3-9}) and
(\ref{eq3-10}), we obtain
\begin{equation}
   \Omega_{\mathrm{gw}}h^2 = \left(\frac{g_0}{g_*}\right)^{1/3}
    \frac{\Omega_Rh^2}{\rho_R(t_i)}\frac{G k^3}{2\pi^2V}I_k, 
    \label{eq3-11}
\end{equation}
where
\begin{equation}
   I_k = \int d\Omega_k\sum_{i,j}
    \left\{\left|\int^{\tau_f}_{\tau_i}d\tau'\cos(k\tau')
    a(\tau')T^{TT}_{ij}(\tau',{\bf k})\right|^2 
    + \left|\int^{\tau_f}_{\tau_i}d\tau'\sin(k\tau')
    a(\tau')T^{TT}_{ij}(\tau',{\bf k})\right|^2 \right\}. 
    \label{eq3-12}
\end{equation}
Here $g_0=3.36$ is the number of radiation degrees of freedom today,
$\Omega_Rh^2=4.15\times 10^{-5}$ is the ratio of the radiation energy
density to the critical energy density today, and $\rho_R(t_i)$ is the
radiation energy density at the initial time. $\Omega_k$ is a unit vector representing 
the direction of ${\bf k}$ and $d\Omega_k=d\cos\theta d\phi$.
We set the scale factor at the initial
time as $a(t_i)=1$. Meanwhile, the frequency observed today is given by
the comoving momentum redshifted by $a(t_0)^{-1}$
\begin{equation}
    f = \frac{k}{2\pi a(t_0)} = \frac{k}{2\pi}
     \left(\frac{g_0}{g_*}\right)^{1/3}\frac{T_0}{T_i}, 
     \label{eq3-13}
\end{equation}
where $T_0=2.725$K is the temperature of the universe observed today and
$T_i$ is the initial temperature (to be specified later).

The computational strategy is as follows: first, we obtain the time
evolution of scalar fields in the real space $\phi(t,{\bf x})$ by the
lattice simulation, then we compute the TT projected stress-energy tensor
as Eqs. (\ref{eq3-3})-(\ref{eq3-5}), and finally we increment the time
integral in Eq. (\ref{eq3-12}) to obtain the spectrum by using
Eqs. (\ref{eq3-11}) and (\ref{eq3-13}).

\section{Numerical simulation}\label{sec4}

\subsection{Problem in Numerical Study and Choice of Parameters}\label{sec4A}

The evolution equation of domain walls is highly nonlinear and we are
forced to use numerical analysis. However, there is a technical problem
in computing the evolution of domain walls numerically: one must
consider two extremely different length scales simultaneously. One is
the width of the wall $\delta_w\sim \eta^{-1}$, which must be bigger
than a lattice spacing throughout the numerical simulation in order to
resolve domain walls by individual unit of lattice.  But if we perform
simulations in the comoving box, the physical lattice spacing expands as
$\propto a(t)$, and becomes unable to resolve
the wall in the finite
simulation time scale. The other length scale is the Hubble horizon,
which must be smaller than the simulation box in order to avoid unphysical
effect caused by the finite size of the simulation box. Unfortunately,
there is a large gap between these two length scales by a factor of the
ratio between the mass scale of the scalar field and the Planck scale as
\begin{equation}
   \delta_w \sim \eta^{-1} \ll \left(\frac{M_P}{\eta}\right)\eta^{-1}
    \sim H^{-1}. 
    \label{eq4-1}
\end{equation}
Therefore, we can not simulate the physical evolution of domain walls
for a long time for energy scale $\eta$ sufficiently smaller than the Planck
scale.

The standard PRS algorithm~\cite{1989ApJ...347..590P} avoids this
difficulty described above by keeping the {\it comoving} wall width
constant in time, to maintain the width resolvable throughout a
simulation. However, in this work we can not apply this method, since
the energy of gravitational waves radiated by domain walls is estimated
incorrectly due to the unphysical nature of the wall width.

Hence we directly solve the physical equation of motion
\begin{equation}
    \ddot{\phi} + 3H\dot{\phi}-\frac{\nabla^2}{a^2}\phi 
     +\frac{dV}{d\phi} = 0, 
     \label{eq4-2}
\end{equation}
with the potential given by Eq.~(\ref{eq2-1}). This restricts the time
scale of the simulation and the model parameters. We must choose $\eta$
close to the Planck scale due to the fact represented by Eq. (\ref{eq4-1}),
but if we choose $\eta$ close enough to $M_P$, the onset of scaling is
delayed in the unit of Hubble time (we choose time coordinate normalized
by the initial Hubble time $t_i \sim H^{-1}(t_i)$, see Appendix
\ref{secA1}) due to the degeneracy between the inverse mass scale
$\eta^{-1}$ and the Hubble radius $H^{-1}$. As the marginal possible
parameters, we choose $\lambda =1.0$ and $\eta=1.05\times
10^{17}$GeV. We assume the radiation dominated era and the relativistic
degree of freedom is constant given by $g_*=100$ during the
simulation. The final time is set to be $t_f=151t_i$ where $t_i$ is
initial time chosen so that the initial temperature is twice of the
critical temperature of the phase transition ($T_i=2T_c=4\eta$,
$t_i=t_c/4$). We performed lattice simulations with $256^3$ points in
the cases with $b=25$ and $b=50$, where $b$ is the (comoving) box size
in the unit of $t_i$. In these parameters, the resolution of the wall
width is well maintained and the horizon scale is smaller than the
simulation box even at the final time (see Appendix \ref{secA1}).

For biased domain walls, we choose the parameter $\epsilon$ such that
the conditions Eqs.~(\ref{eq2-9}) and (\ref{eq2-12}) are satisfied:
\begin{equation}
   \frac{16\pi}{9}\lambda\left(\frac{\eta}{M_P}\right)^2 
    < \epsilon < 4.7\times (\lambda g_*)^{1/2}\frac{\eta}{M_P}. 
    \label{eq4-3}
\end{equation}
In the parameters described above, Eq.~(\ref{eq4-3}) corresponds to the
condition $4.1\times 10^{-4}<\epsilon<0.40$. We run the simulations
varying $\epsilon$ in the range $\epsilon =$ 0.015, 0.02, 0.025, 0.03,
0.035, and 0.04 which satisfy the condition Eq.~(\ref{eq4-3}). These
values also satisfy the condition for an appearance of the infinite size
of domains, Eq.~(\ref{eq2-7}).

We emphasize that the dynamical range of the simulations is rather small compared with realistic cosmological dynamical ranges. The actual dynamical range is 12 in conformal time. For the domain wall evolution, this is reduced to less than $(151t_i/30t_i)^{1/2}\simeq$ 2.2 since the result with $\epsilon=0$ do not approach scaling until $t=30t_i$ (see next subsection and Fig.~\ref{fig2}). On the other hand, the typical time scale for the wall domination is 64 [according to Eq.~(\ref{eq2-10})]. It is extremely difficult to improve dynamical ranges without using PRS algorithm. Future simulations with much larger dynamical ranges should confirm and improve the results of our current simulations.

\subsection{\label{sec4B}Evolution of Domain Wall Networks}

\begin{figure}[htbp]
\begin{center}
\setlength{\tabcolsep}{20pt}
\begin{tabular}{c c}
\resizebox{50mm}{!}{\includegraphics{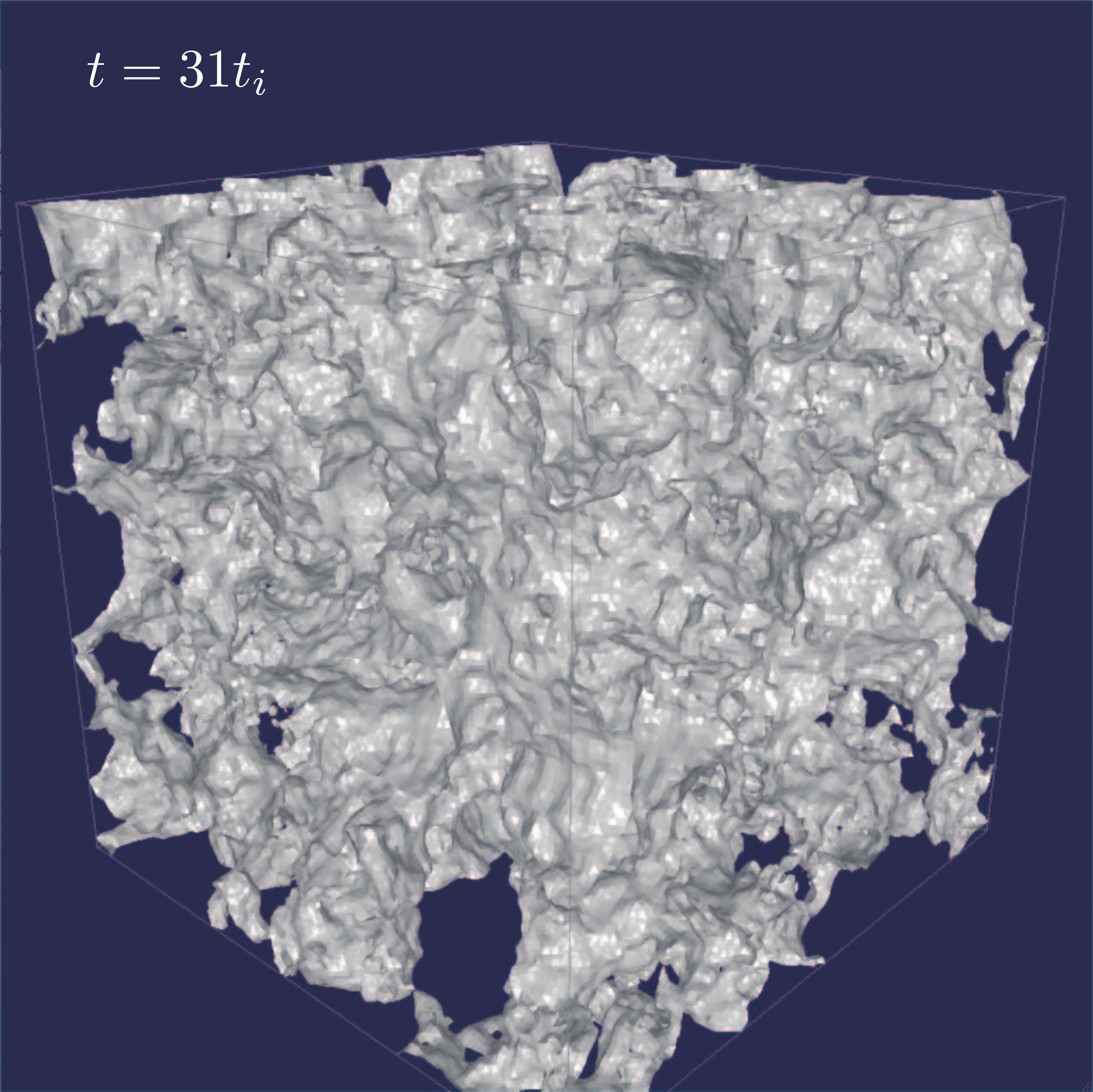}} &
\resizebox{50mm}{!}{\includegraphics{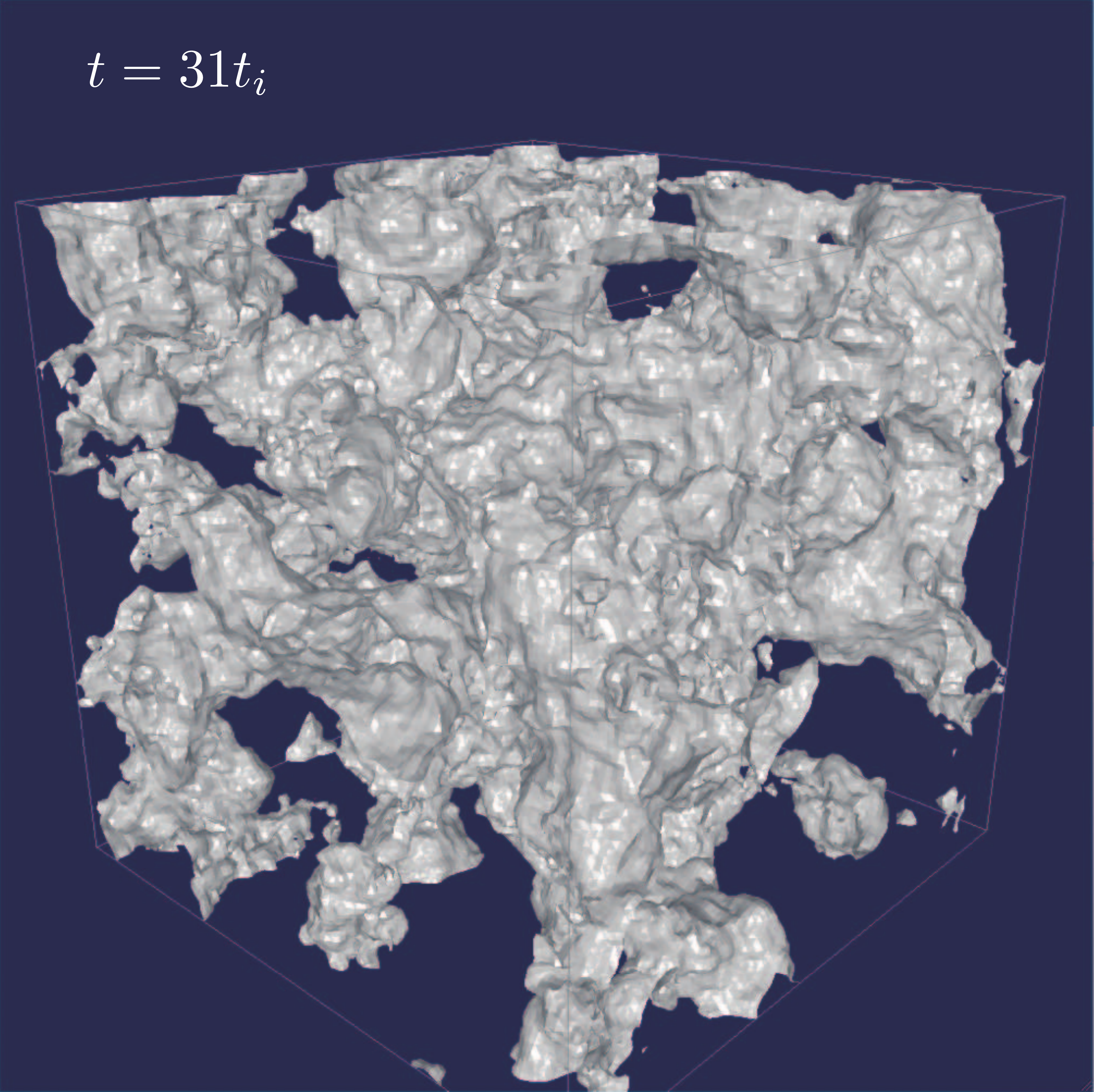}} \\
\resizebox{50mm}{!}{\includegraphics{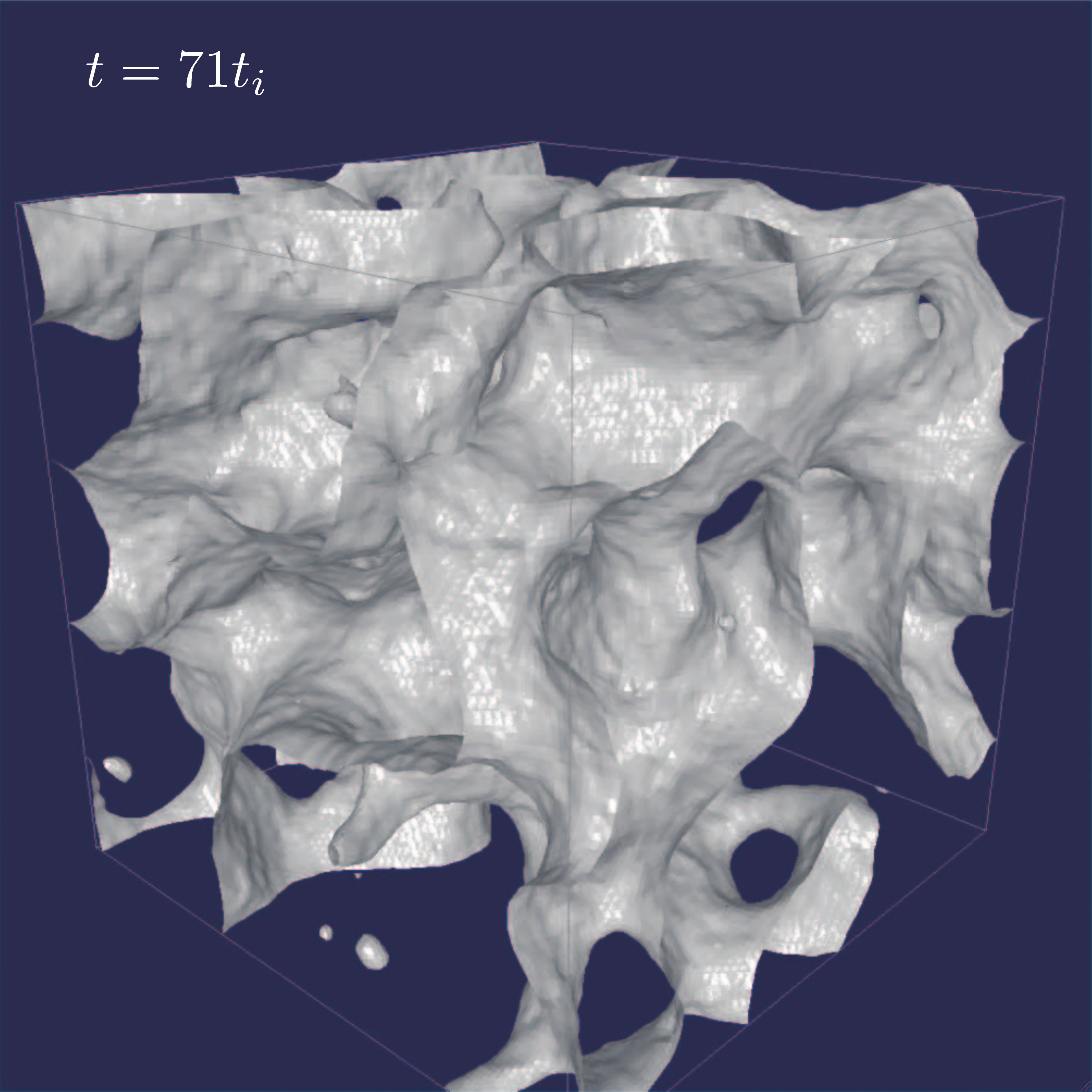}} &
\resizebox{50mm}{!}{\includegraphics{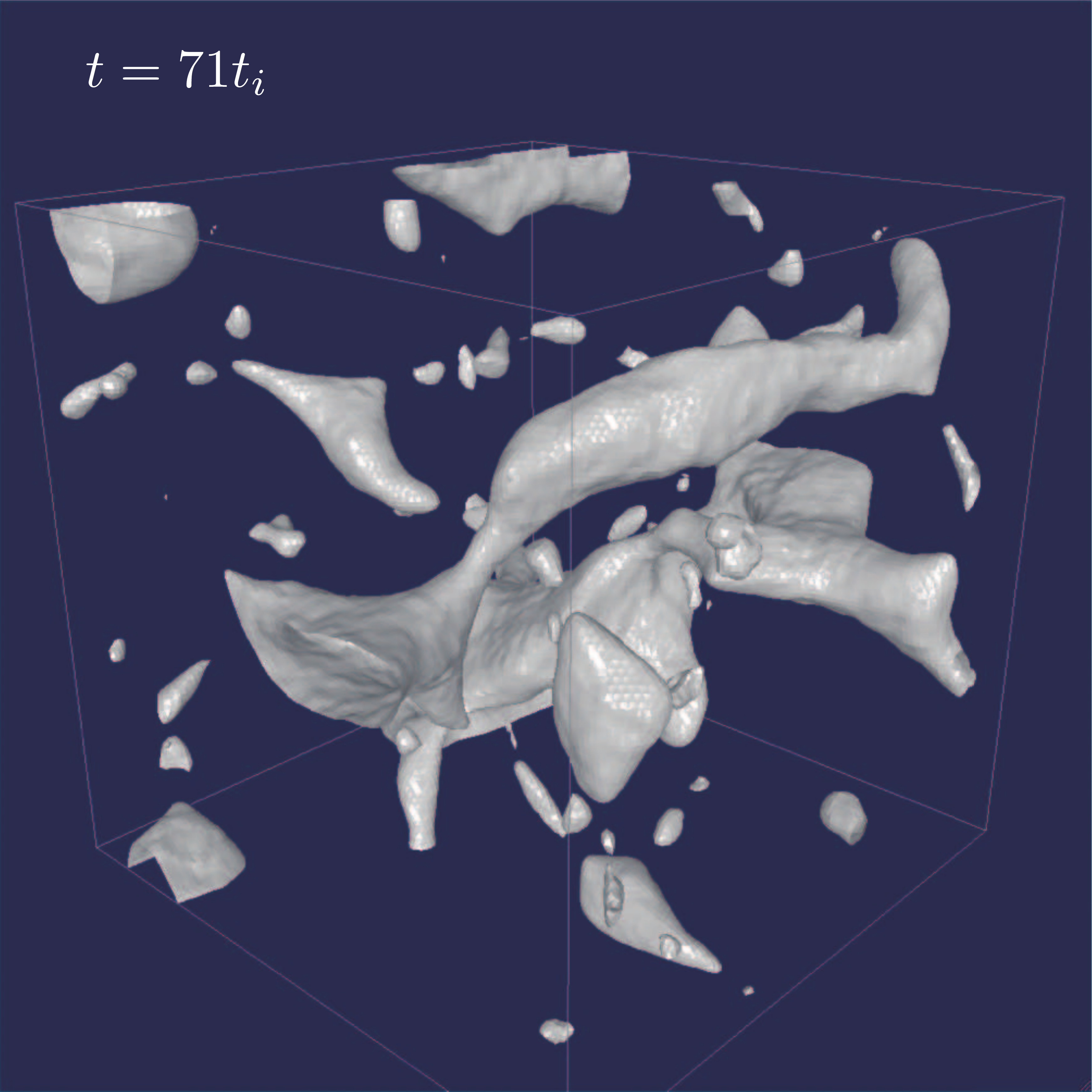}} \\
\resizebox{50mm}{!}{\includegraphics{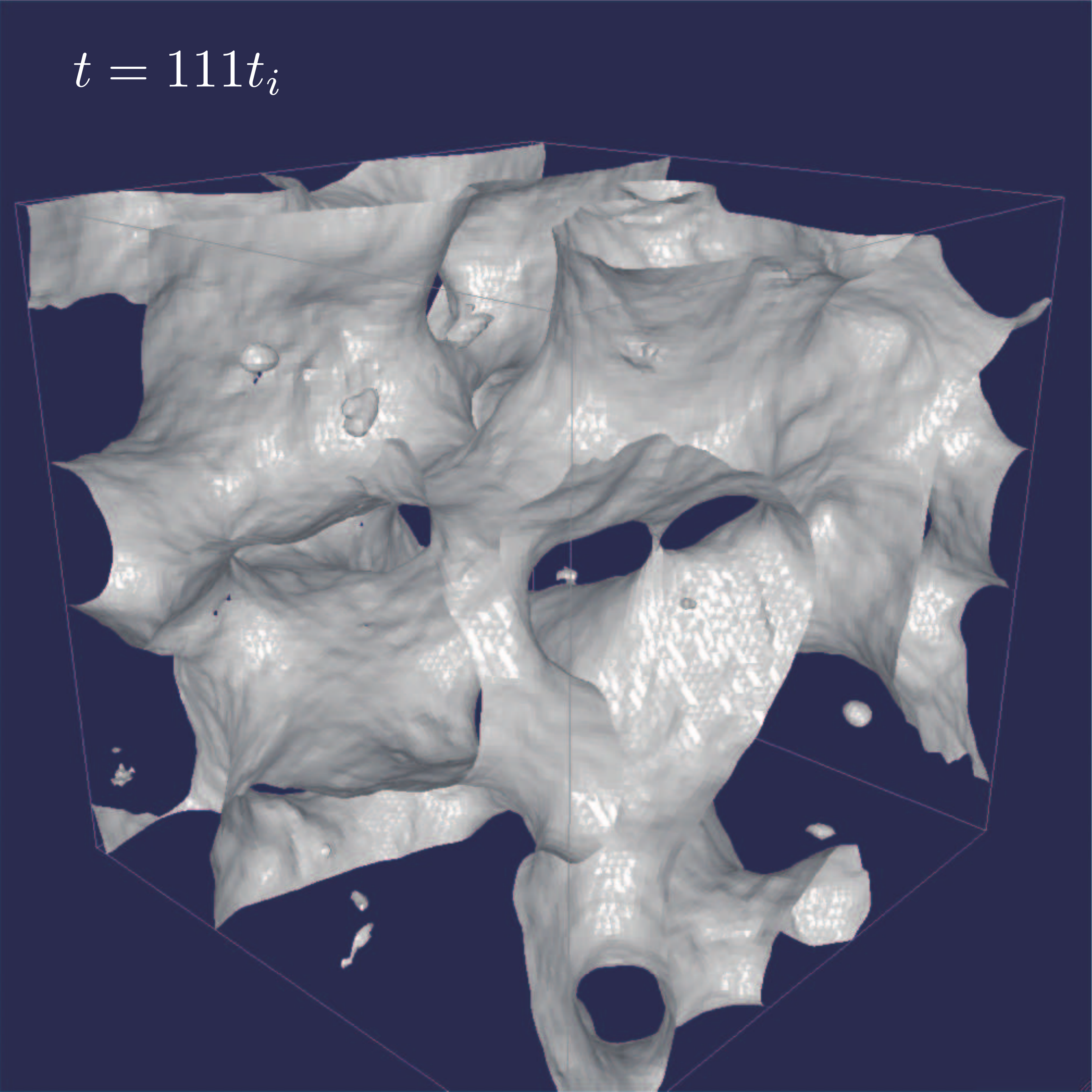}} &
\resizebox{50mm}{!}{\includegraphics{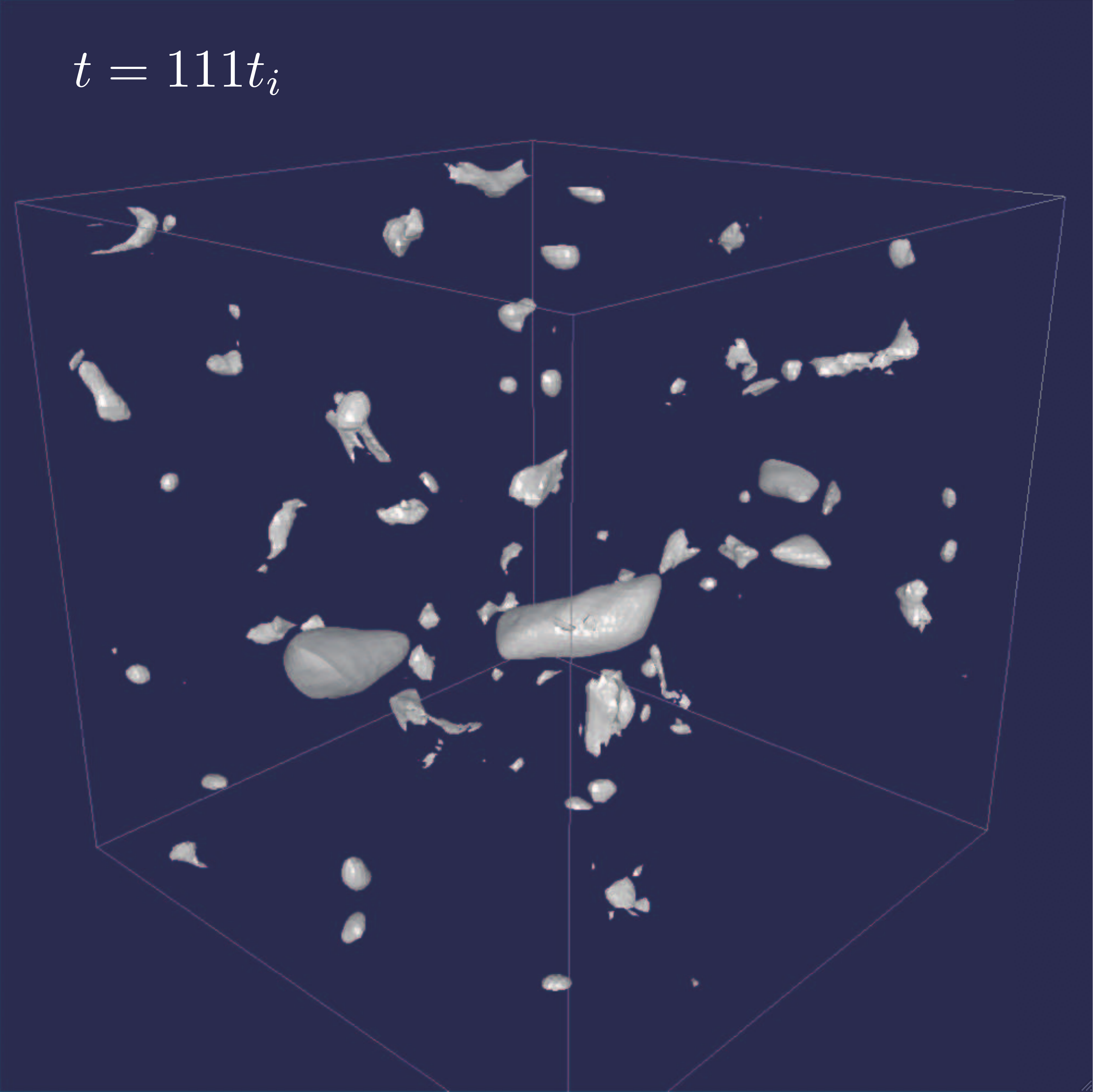}} \\
\end{tabular}
\end{center}
\caption{The evolution of stable ($\epsilon=0$: left) and unstable
($\epsilon=0.02$: right) domain walls. The white surface corresponds to
the region where the value of $\phi$ crosses the zero.}  
\label{fig1}
\end{figure}

The typical evolution of stable ($\epsilon=0$) and biased ($\epsilon\ne
0 $) domain wall networks is shown in Fig.~\ref{fig1}. The initial
thermal fluctuations of the scalar field are damped around $t=30t_i$,
and domain walls extended over the horizon scale are formed. In the case
$\epsilon=0$, the wall networks evolve with annihilation and
reconnection to maintain scaling property. If we turn on the bias,
instead, the region occupied by false vacuum after the phase transition
is smaller than the unbiased case, and walls gradually collapse due to
the volume pressure.

To see more quantitatively, we calculate time evolution of the comoving
area density of domain walls as shown in Fig.~\ref{fig2} (see Appendix
\ref{secA3} for a detail of the calculation). From this figure we can
see that the scaling regime in which the area density scales as
$t^{-1/2}$ begins around $t=30t_i$ for the case $\epsilon=0$. In the
$b=25$ case, the scaling law breaks down around the late stage of the
evolution. This may be caused by an unphysical effect of the
finiteness of the simulation box, that is, the horizon scale become
comparable to the size of the simulation box around the final time (the
similar deviation from the scaling solution can be found in
\cite{1989ApJ...347..590P,2005PhRvD..71h3509O,2005PhLB..610....1A}).

\begin{figure}[htbp]
\begin{center}
\begin{tabular}{c c}
\resizebox{90mm}{!}{\includegraphics{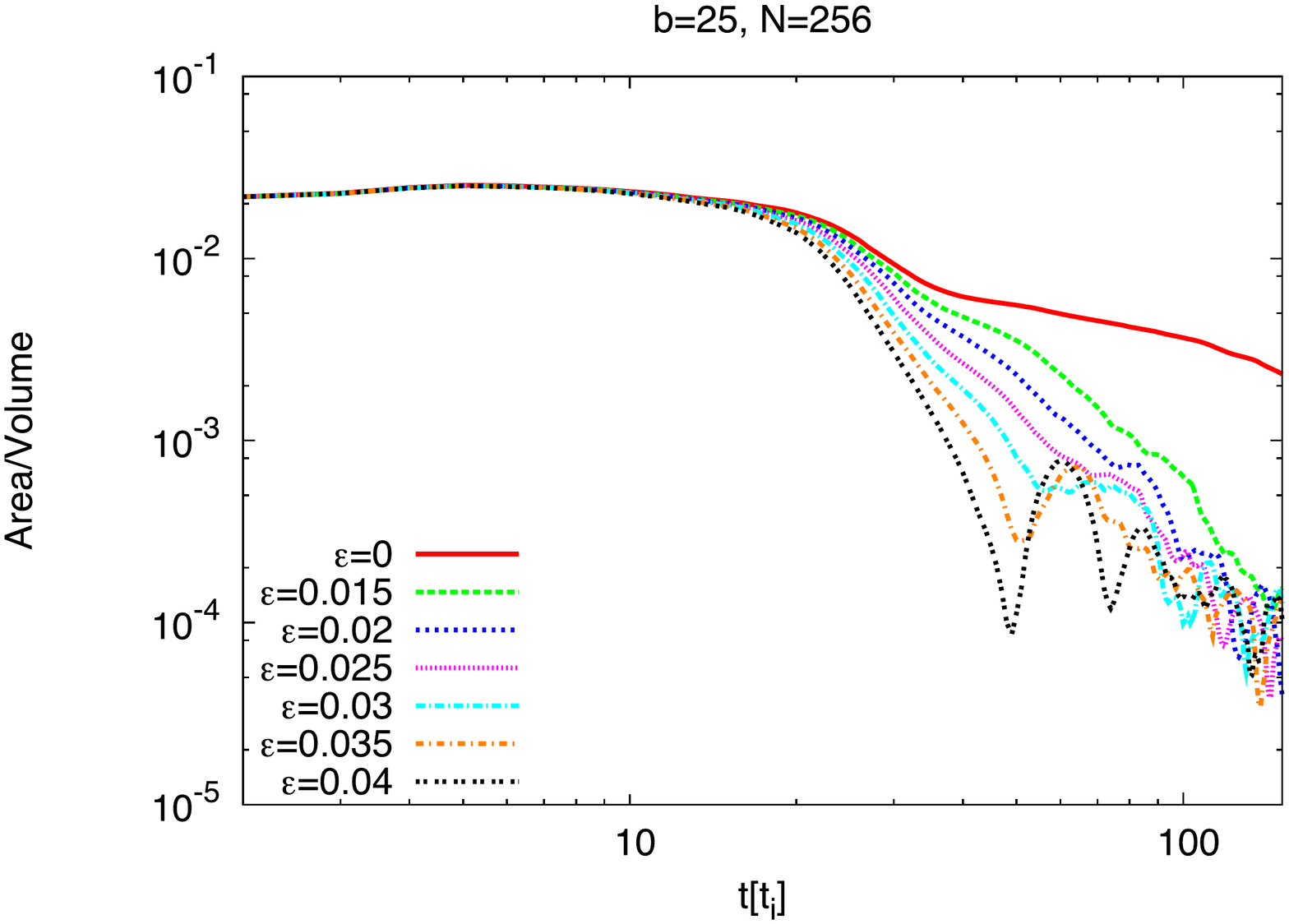}} &
\resizebox{90mm}{!}{\includegraphics{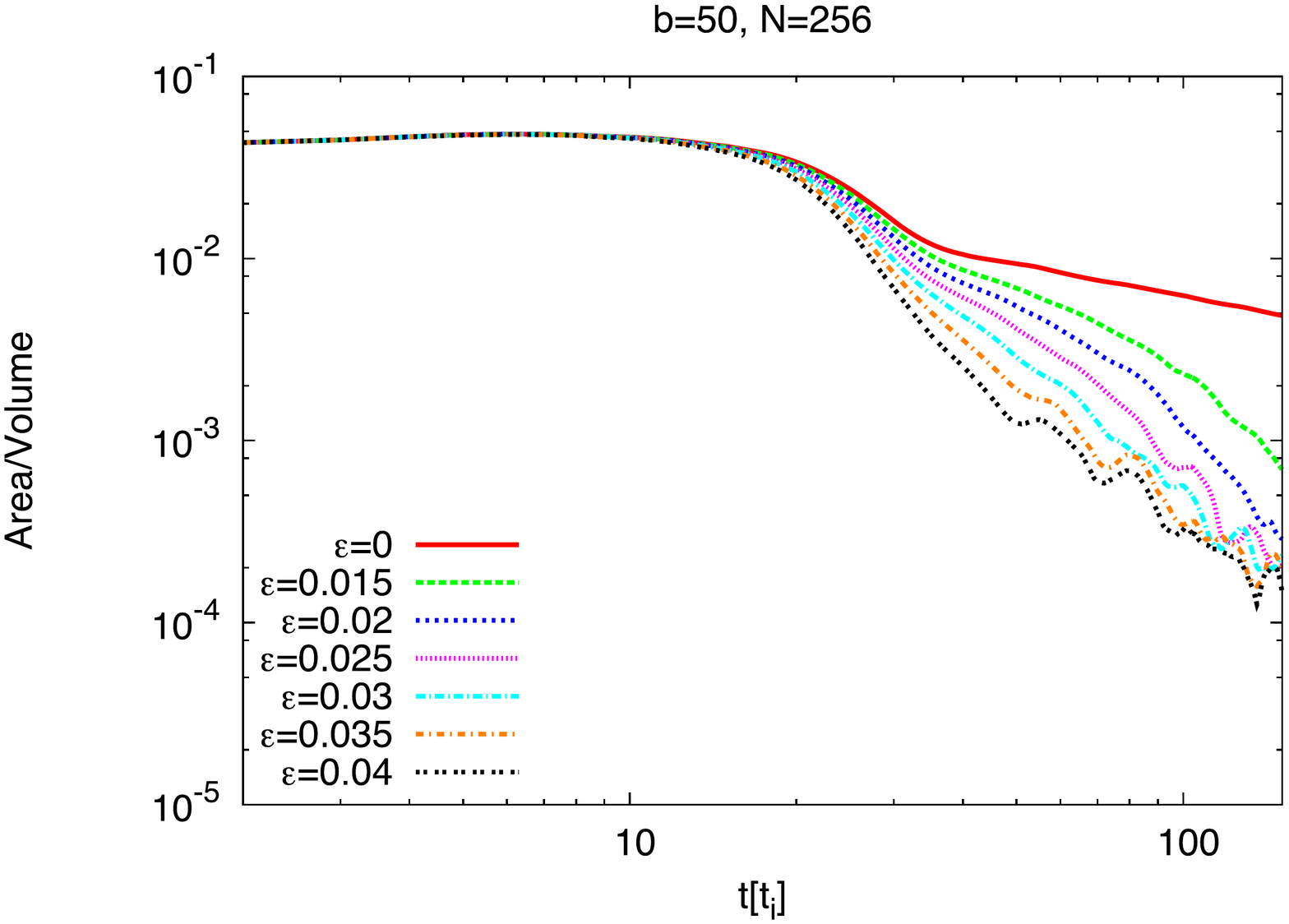}} 
\end{tabular}
\end{center}
\caption{The time evolution of the comoving area density of domain walls
for various values of $\epsilon$. The left panel shows the result in the
case with the box size $b=25$ and the right panel shows that in the case
with $b=50$.}
\label{fig2}
\end{figure}

In the primary thermal stage $t \lesssim 30t_i$, the correlation length
of the scalar field is determined by mass scale of the scalar field
$\xi_{\mathrm{corr}}\sim m^{-1} \sim \eta^{-1}$ rather than the Hubble
horizon scale. In this stage, the value of the scalar field changes
randomly in the short length scale given by $\xi_{\mathrm{corr}}$, thus
$A/V$ remains nearly constant. Note that if $bt_i > \xi_{\mathrm{corr}}
\sim m^{-1}$, by increasing box size $b$ with the fixed number of grids
one can count more points where the scalar field change their sign, leading
to increase the value of $A/V$. In fact, the value of $A/V$ at the
primary stage $t \lesssim 30t_i$ in the case $b=50$ is larger than that
in the case $b=25$, as shown in Fig.~\ref{fig2}.

As we turn on the bias, the value of $A/V$ falls by a factor of ${\cal
O}$(10-100) in the simulation time scale. The typical time scale of the
decay of the area density is well described by the estimation
Eq. (\ref{eq2-11}) except a factor of ${\cal O}$(1). There are
oscillations of $A/V$ after the decay of domain walls. This can be
interpreted as the radiation of the energy stored in walls as the scalar
field oscillating around the true vacuum \cite{1997PhRvD..55.5129L}. 
The difference in the amplitude of oscillations between the case with $b=25$
and $b=50$ may be due to the poor resolution in the simulation with $b=50$.
If the wavelength of the scalar field radiations produced by the decay of domain walls 
is comparable or smaller than the small scale cutoff ($\sim\eta^{-1}$), the typical length 
scale over which the scalar field varies becomes shorter than the lattice spacing. 
This makes the value of $A/V$ inaccurate because we estimate the value $A/V$ as a sum of
the link of points where $\phi$ has different signs (see Appendix~\ref{secA3}). 
In the case with $b=50$, the lattice spacing $\delta x$ is larger than that with $b=25$ and
it is likely to lose many links over which the scalar field changes its signs. 
Therefore, the oscillation of the $A/V$ in the case with $b=50$ is less violent than that with $b=25$, as we see in Fig.~\ref{fig2}.
\subsection{Spectrum of Gravitational Waves}
\label{sec4C}

\begin{figure}[htbp]
\begin{center}
\setlength{\tabcolsep}{10pt}
\begin{tabular}{c c}
\resizebox{80mm}{!}{\includegraphics[angle=-90]{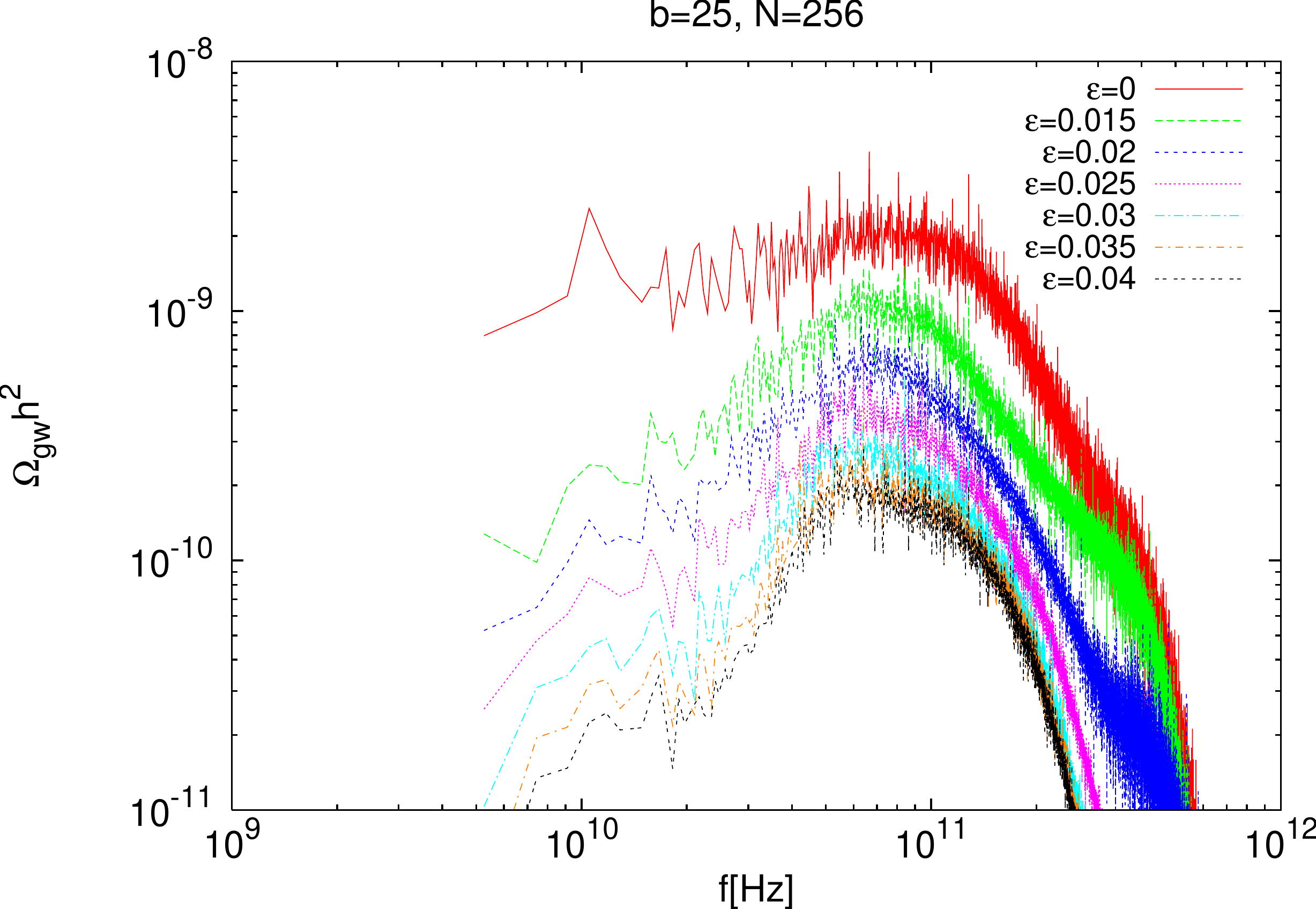}} &
\resizebox{80mm}{!}{\includegraphics[angle=-90]{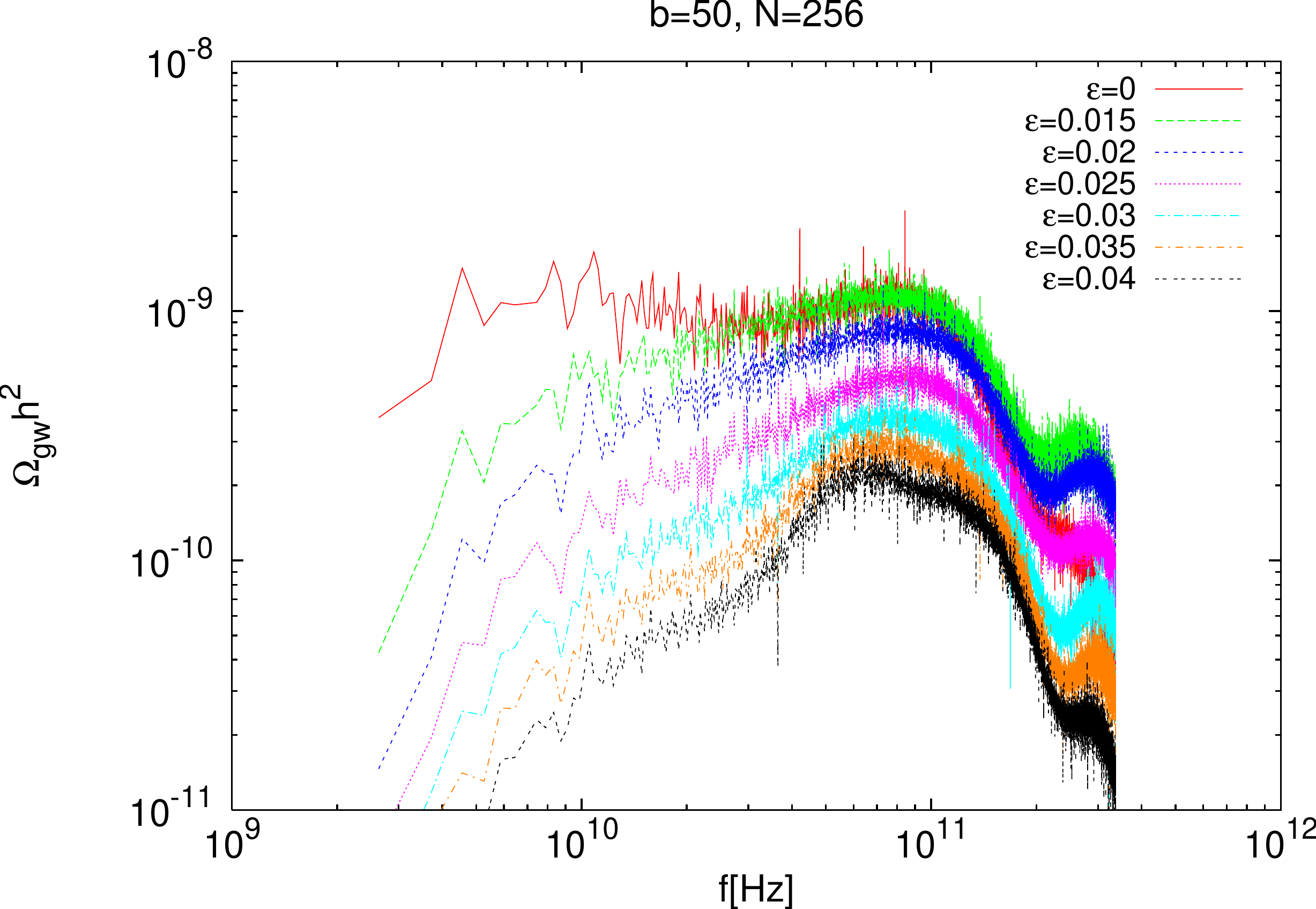}} 
\end{tabular}
\end{center}
\caption{The spectrum of gravitational waves from domain walls for
various values of $\epsilon$. The left panel shows the result in the
case with the box size $b=25$ and the right panel shows that in the case
with $b=50$.}  
\label{fig3}
\end{figure}

By using the method described in Section~\ref{sec3}, we compute the
spectrum of gravitational waves produced by domain walls. The results are
shown in Fig.~\ref{fig3}. To take small $b$ with fixed number of lattice
points corresponds to increasing the spatial resolution. Therefore, the
frequency range computed in the case with $b=25$ is higher than that in
the case with $b=50$. The two figures shown in Fig.~\ref{fig3} seem to
be different in spite of the fact that the two results are obtained from
calculations with completely the same parameters except $b$. For
example, there is a peak in high frequency edge around
$2\times10^{11}$Hz of the spectrum with $b=50$, but the corresponding
peak is not found in the spectrum with $b=25$. Hence this high frequency
peak in $b=50$ is thought to be unphysical noise due to the poor
resolution for a small scale. Another difference is that the low
frequency tail of the spectrum in $b=25$ is steeper than that in
$b=50$. This is due to the finite volume effect which promotes walls to
collapse around the final time in $b=25$, as shown in
Fig. \ref{fig2}. Because of the limitation in our computer memory, we
can not perform the simulation with maintaining physical resolution over
full range of gravitational wave spectra. From now on, therefore, we use
the result of the case with $b=50$ and ignore the high frequency peak
around $2\times 10^{11}$Hz.

\begin{figure}[htbp]
\begin{center}
\begin{tabular}{c c}
\resizebox{80mm}{!}{\includegraphics[angle=-90]{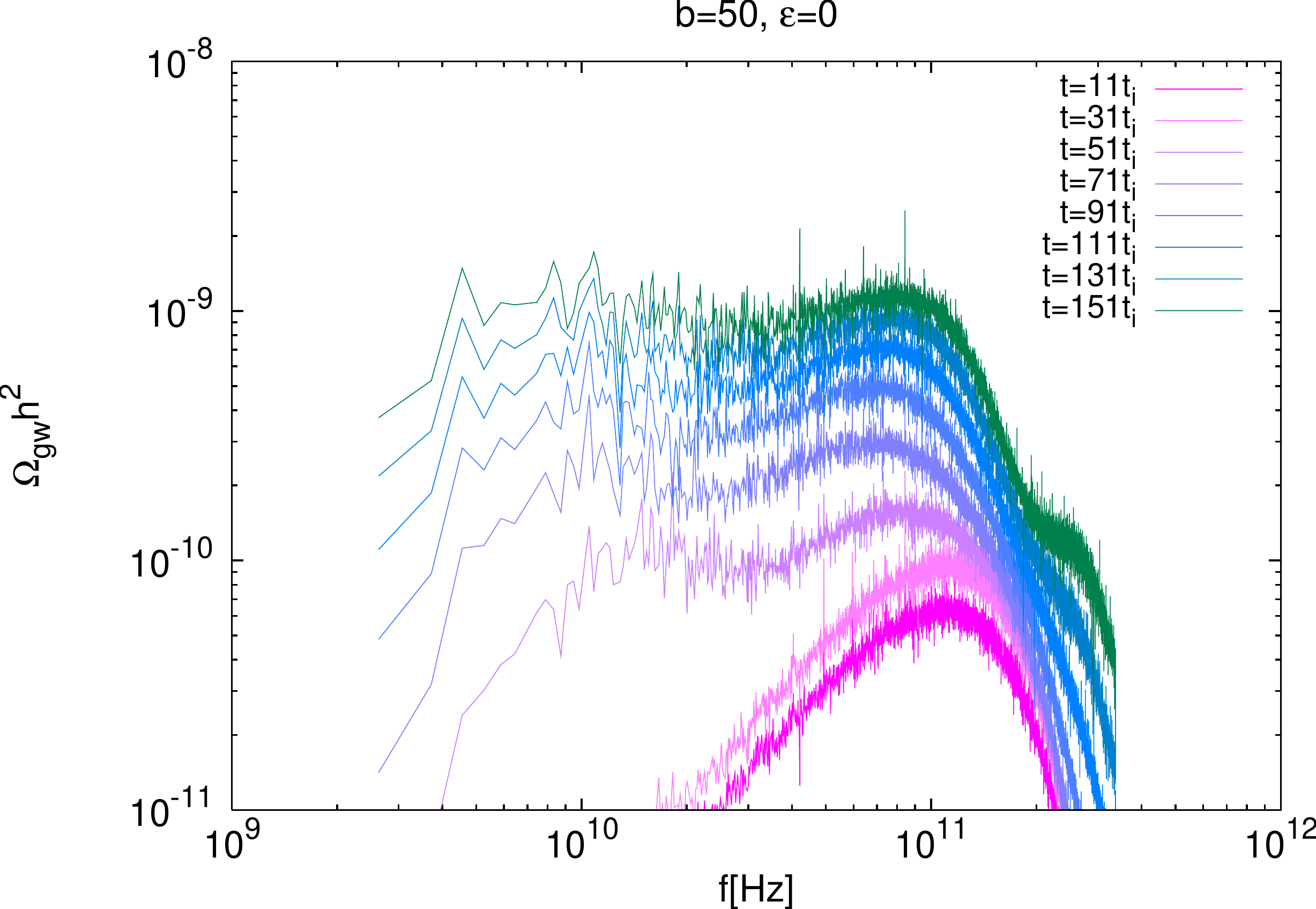}} &
\resizebox{80mm}{!}{\includegraphics[angle=-90]{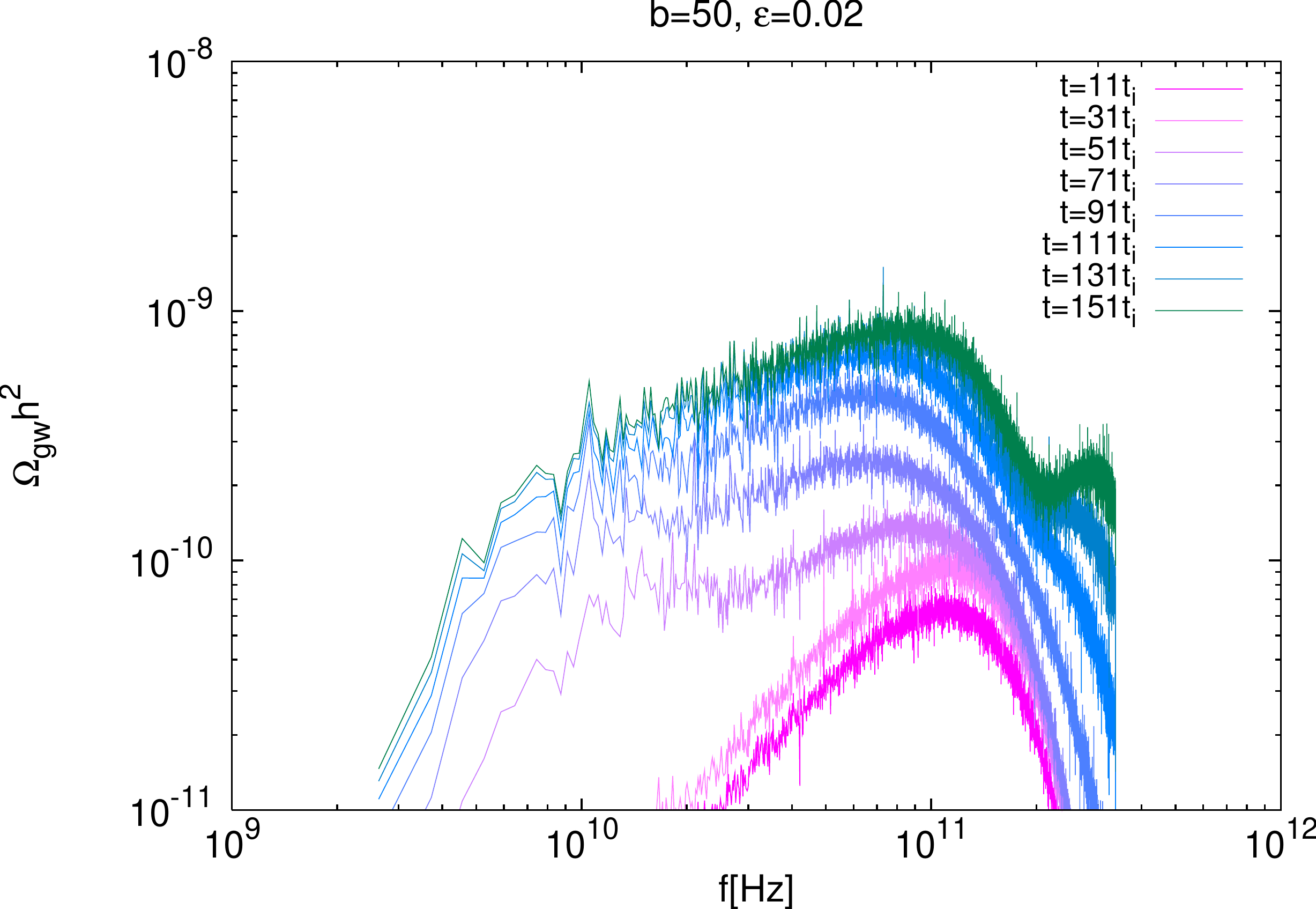}} \\
\resizebox{80mm}{!}{\includegraphics[angle=-90]{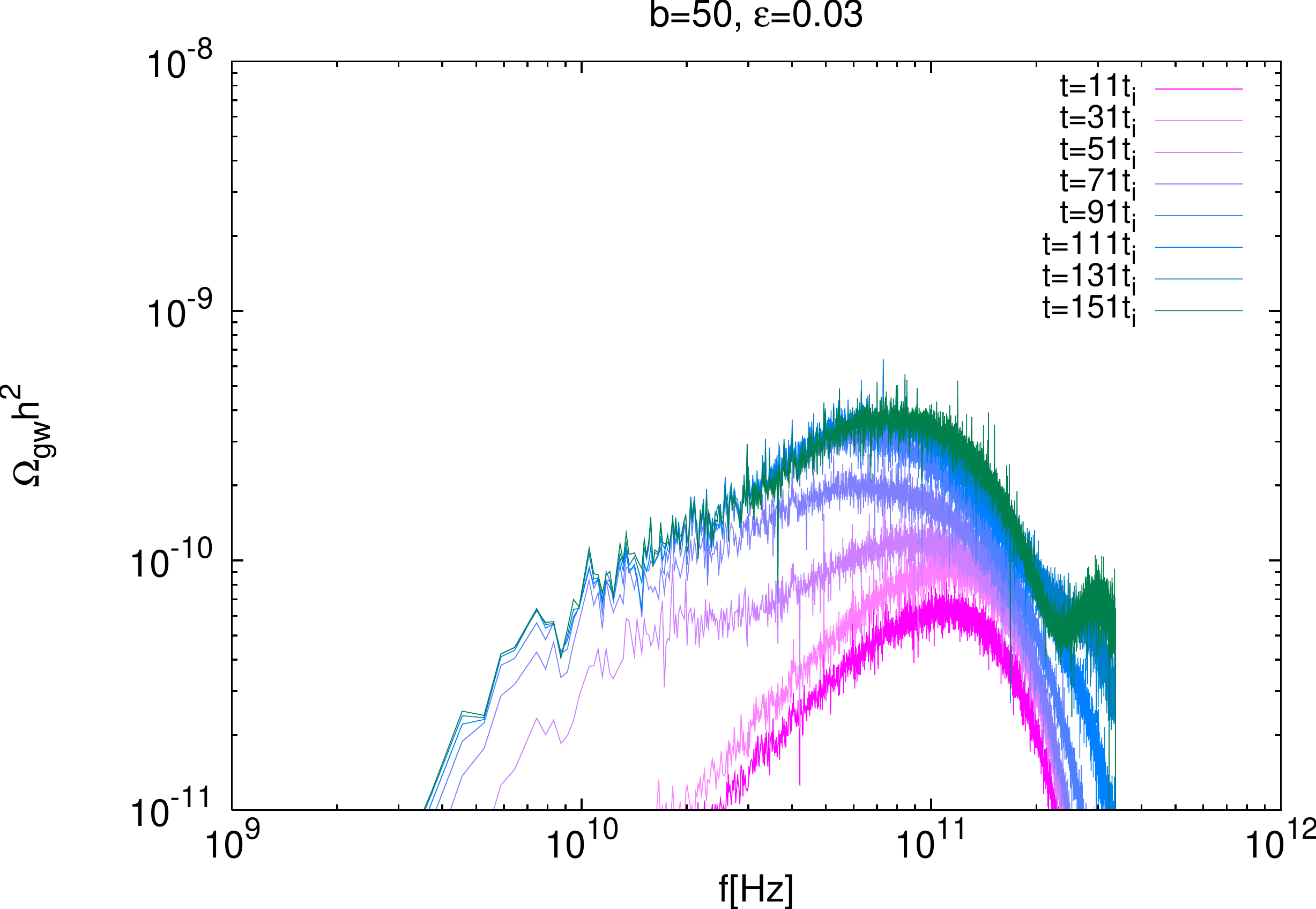}} &
\resizebox{80mm}{!}{\includegraphics[angle=-90]{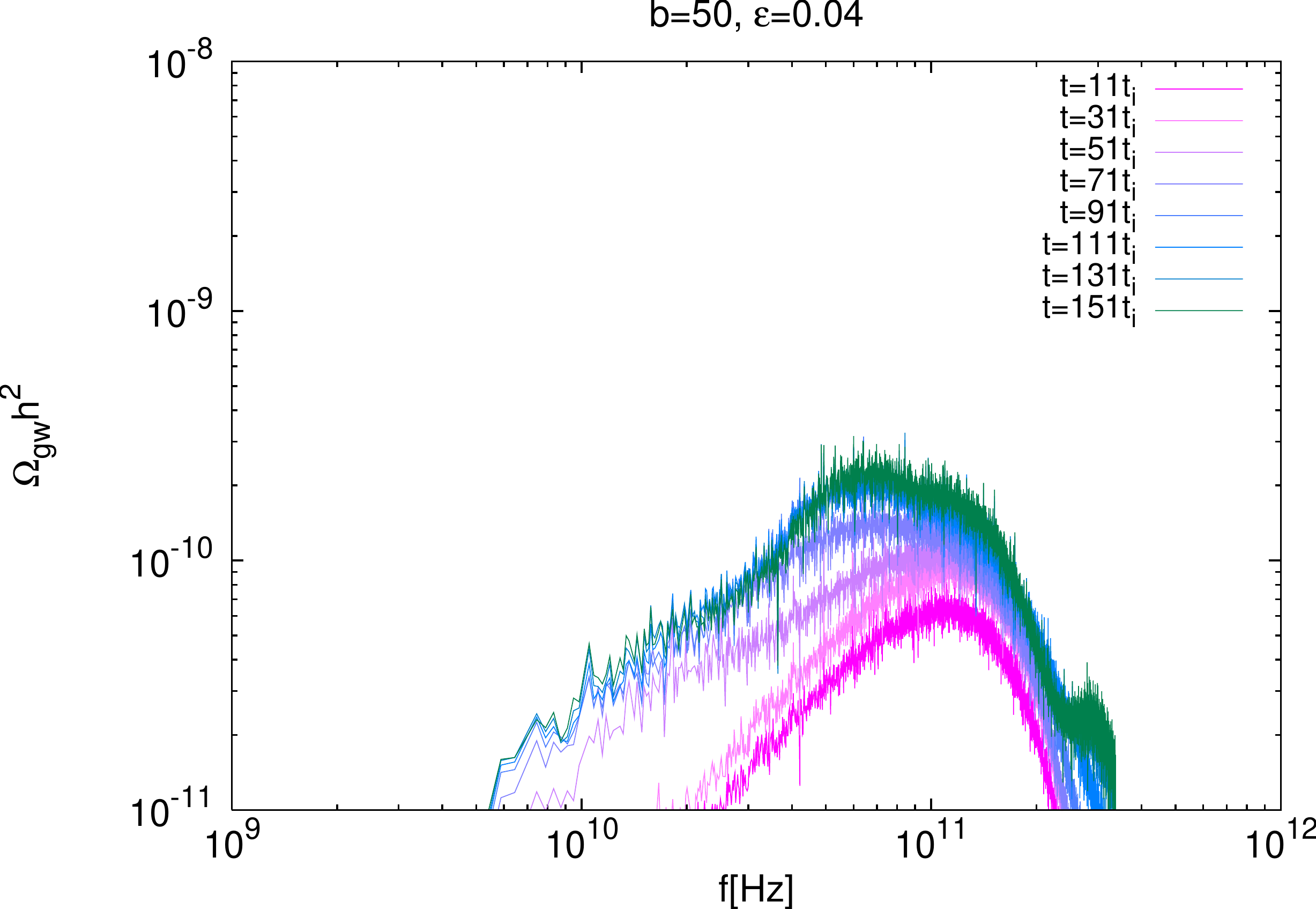}} \\
\end{tabular}
\end{center}
\caption{The time evolution of the gravitational wave spectra for
various values of $\epsilon$ (in the case $b=50$). The spectra are shown
from the time $t=11t_i$ (pink) to $t=151t_i$ (green) with the interval
$\Delta t=20t_i$.}  
\label{fig4}
\end{figure}

In Fig.~\ref{fig4}, we show the time evolution of the gravitational wave
spectra for several values of $\epsilon$ in the case $b=50$. There are also the
unphysical high frequency peak in those spectra, but we ignore them for
the reason described above. By comparing Fig.~\ref{fig2} and
Fig.~\ref{fig4} we can see that the spectrum of gravitational waves
evolves correspondingly to the evolution of the scalar field. In the
primary stage ($t\lesssim30t_i$), the thermally distributed scalar field
produces the spectrum which has a peak around the frequency
$f\sim10^{11}$Hz, and after the formation of scaling wall networks
($t\gtrsim 50t_i$) there is an increase in gravitational wave amplitudes
of low frequency modes. These low frequency modes turn out to correspond
the horizon scale at the time when gravitational waves are produced. We will
investigate it in detail in Section \ref{sec5}. In the late stage, if
$\epsilon=0$, the amplitudes of the gravitational waves keep growing in both
high frequency of the small scale and low frequency of the horizon
scale. On the other hand, if $\epsilon\ne0$, the growth in low frequency
terminates when walls begin to collapse, and only high frequency modes
grow in the late stage. The resulting spectrum has the low frequency
edge and the high frequency peak, and gradually increases in
intermediate modes.

\section{Analytic estimation and perspective for
 observations}
\label{sec5}

In this section we discuss the physical interpretation of the numerical
results. We check whether the results of the numerical simulation match
the naive analytic estimations. Finally we extrapolate the results into
the general parameter space and estimate the amplitude and the frequency
of gravitational waves observed today for various parameters. In
particular we discuss whether the spectrum is observable in the future
gravitational wave experiments.

\subsection{Physical Interpretation of the Spectrum}
\label{sec5A}

There are two mechanisms relevant for production of gravitational waves
from domain wall dynamics. One is due to the relativistic motion of
walls in the universe, and another is due to the interactions of domain
wall networks such as collision, separation, and reconnection of
walls. Considering the scaling property of domain walls, the typical
scale relevant for the former is the horizon scale at the time
when gravitational waves are produced.
Meanwhile, the typical scale relevant for the latter is the small scale
of the structure on the domain wall. As long as domain walls are in the
scaling regime, both of these two mechanisms are efficient, and the
resulting spectrum extends over a broad frequency band between the low
frequency of the horizon scale and the high frequency of the small scale
structure of the wall, as shown in Fig.~\ref{fig3}. Instead, the growth
of the amplitude in low frequency modes terminates when domain walls
begin to collapse if there is a bias.

From the above considerations, we conjecture that the form of the spectra is
determined as follows. There is a peak in the high frequency
corresponding to the small scale structure of the wall (i.e. the wall
width). The spectrum gradually decreases toward the low frequency, and
there is an `edge' in the low frequency corresponding to the horizon
scale. For causality reasons, there are no modes beyond the horizon
scale.

In the following, we estimate the amplitude and the frequency of
gravitational waves analytically in order to justify the above
considerations.

\subsection{Analytic Estimation}
\label{sec5B}

First we estimate the peak frequency. The peak is given by the
interaction of domain wall networks and we naively estimate that the
peak frequency corresponds to the width of the domain wall. One can
think of this process as the collision of wave packets of classical
scalar fields whose wavelength is about $\lambda_{\mathrm{peak}}\sim
\delta_w \sim (\lambda^{1/2}\eta)^{-1}$. Then, the physical wave number
of the peak of gravitational waves at the time $t_*$ when they are
produced is
$k_{\mathrm{peak}}/a(t_*)=2\pi/\lambda_{\mathrm{peak}}=2\pi\lambda^{1/2}\eta$,
where $k_{\mathrm{peak}}$ is the comoving wave number corresponding to
the peak frequency. Taking redshift into account, the peak frequency
observed today is obtained as
\begin{eqnarray}
    f_{\mathrm{peak}} &=& \frac{k_{\mathrm{peak}}}{2\pi a(t_0)} 
     = \left(\frac{a(t_*)}{a(t_0)}\right)\lambda^{1/2}\eta 
     \nonumber\\
     &=& 3.5\times
      10^{11}\times\lambda^{1/2}\left(\frac{100}{g_*}\right)^{1/3}
      \left(\frac{t_*}{151t_i}\right)^{1/2} \mathrm{Hz}. 
      \label{eq5-1}
\end{eqnarray}
Setting $t_*$ to the final time of the simulation, $151t_i$, the
estimation of the peak frequency gives $3.5\times 10^{11}$Hz, which
agrees with our numerical results except the factor of ${\cal
O}$(1). Note that if we assume the entropy conservation, the dependence
$k_{\mathrm{peak}}\propto\eta$ is canceled by the factor
$a(t_i)/a(t_0)\propto 1/\eta$. Therefore, generically the peak frequency
is located in the high frequency of order $10^{11}$Hz regardless of the
value of $\eta$.

Next, we estimate the amplitude of gravitational waves observed
today. The energy of gravitational waves produced by domain walls is estimated
as $E_{\mathrm{gw}} \sim GM_{DW}^2/R_*$, where $R_*$ is a typical length
scale over which the energy distributes and $M_{DW}$ is the mass scale
of the domain wall. 
We do not specify the value of $R_*$ since the result is not depend on $R_*$ as shown shortly.
We estimate $M_{DW} \sim \sigma R_*^2 =
\sqrt{2\lambda}\eta^3R_*^2/3$, and the energy density of gravitational
waves at the production time is $\rho_{\mathrm{gw}}(t_*)\sim
E_{\mathrm{gw}}/R_*^3\sim (8\lambda/9)(\eta/M_P)^2\eta^4$. By
redshifting it, we find the amplitude of the gravitational waves observed
today as
\begin{equation}
   \Omega_{\mathrm{gw}}h^2 \sim 
    \left(\frac{a(t_*)}{a(t_0)}\right)^4
    \frac{\rho_{\mathrm{gw}}(t_*)h^2}{\rho_{c,0}}
    \sim 7.5\times
    10^{-10}\times\lambda\left(\frac{100}{g_*}\right)^{4/3}
    \left(\frac{\eta}{1.05\times 10^{17}\mathrm{GeV}}\right)^2
    \left(\frac{t_*}{151t_i}\right)^2, 
    \label{eq5-2}
\end{equation}
which agrees with the peak amplitude $\Omega_{\mathrm{gw}}h^2\sim{\cal
O}(10^{-9})$ of the numerical results. 

We also estimate the location of the low frequency edge in the spectrum
which corresponds to the horizon scale. The physical wave number
corresponding to the horizon scale at time $t$ is given by
$k_{\mathrm{hor}}/a(t)=2\pi H(t)=\pi/t$. Hence the frequency observed
today is given by
\begin{equation}
    f_{\mathrm{hor}} =\frac{k_{\mathrm{hor}}}{2\pi a(t_0)} 
     = \left(\frac{a(t)}{a(t_0)}\right)\frac{1}{2t} 
     = 6.58\times 10^{10} \times (t/t_i)^{-1/2}
    \left(\frac{\eta}{1.05\times 10^{17}\mathrm{GeV}}\right)
     \mathrm{Hz}. 
     \label{eq5-3}
\end{equation}
Notice that $t_i \propto \eta^{-2}$ and hence $f_{\mathrm{hor}}$ is
independent of both $\eta$ and $t_i$ as it should be.  For example, the
low frequency edge at the final time of the simulation $t=151t_i$
becomes $f_{\mathrm{hor}}\simeq 5\times 10^9$Hz, which reproduces the
location of the edge in the spectrum obtained by numerical simulations.

Before moving to the next subsection, we discuss the origin of the
spectrum produced during the primary stage ($t\lesssim 30t_i$) of the
scalar field evolution. In this stage, the scalar field fluctuates
randomly with correlation length roughly given by inverse of the
temperature $\xi_{\mathrm{corr}}\sim T^{-1}$, and gravitational waves
can be produced via the quadratic interaction of the scalar field. We
denote the amplitude of the scalar field fluctuations by
$\delta\phi$. From the equation of motion for the metric perturbation,
Eq. (\ref{eq3-2}), a typical amplitude of gravitational wave $h_g$ with
the comoving wave number $k$ is estimated as
$k^2h_g\sim16\pi\nabla\phi\nabla\phi/M_P^2\sim 16\pi
k^2\delta\phi^2/M_P^2$, thus we obtain
$h_g\sim50\times\delta\phi^2/M_P^2$. On the other hand, the ratio
between the energy density of gravitational waves and the total energy
density of the universe at the production time is estimated as
$(\Omega_{\mathrm{gw}})_*\sim\rho_{\mathrm{gw}}/\rho_R(t_*)\sim
\dot{h}_g^2/G\rho_R(t_*)\sim \dot{h}_g^2/H^2(t_*)\sim
t^2_*\dot{h}_g^2\sim h_g^2$, where $\rho_R(t_*)$ is the energy density
of the radiation at $t=t_*$ and we assume the radiation dominated
universe. In our simulations, the scalar field initially fluctuates with
amplitude $\delta\phi\sim T_i \sim \eta \sim10^{17}$GeV, which gives
$(\Omega_{\mathrm{gw}})_*\sim h_g^2 \sim 10^3\times
(\delta\phi/M_P)^4\sim 10^{-5}$. The amplitude of these gravitational waves
observed today is~\cite{1994PhRvD..49.2837K}
\begin{equation}
   \Omega_{\mathrm{gw}}h^2 = 1.67\times 10^{-5}
    \left(\frac{100}{g_*}\right)^{1/3}(\Omega_{\mathrm{gw}})_* 
    \sim 10^{-10}. 
    \label{eq5-4}
\end{equation}
This agrees with our numerical results (see the spectra of $t=11t_i$ and
$31t_i$ in Fig. \ref{fig4}). The location of the peak (i.e. $f\sim
10^{11}$Hz) is almost unchanged from the early time to the final
time. This is because the typical frequency of gravitational waves
produced in the early time is roughly given by correlation length of the
scalar field which is determined by the mass scale $m\sim
\sqrt{3\lambda}\eta$, and this scale is nearly equal to the typical
scale of the small structure on the domain wall $\sim
\lambda^{1/2}\eta$.

\subsection{Fitting the Numerical Results}
\label{sec5C}

The numerical results reveal that the gravitational wave spectra appear
in rather high frequency band. This is partially due to the fact that we
assume very high energy scale, $\eta \simeq 10^{17}$GeV. For the case
with lower $\eta$, it is impossible to perform the numerical computation
correctly because of the technical restriction described in Section
\ref{sec4A}. Instead, we assume the parameter dependence of analytic
estimations obtained above, and fit the numerical factor into the
results of numerical simulations to derive the formulae for the
frequency and the amplitude of peak and edge in the gravitational wave
spectra. In deriving these formulae, 
we check the $\epsilon$ dependence of the numerical results. The
difference in the bias $\epsilon$ affects the time scale of the collapse
of the domain wall networks [see Eq. (\ref{eq2-11})]. In the following
we assume that the typical time of the production of gravitational
waves $t_*$ is identical to the decay time of the domain walls
$t_{\mathrm{dec}}$, and estimate $\epsilon$ dependence of the
spectrum. Then we see if our numerical results follow the dependence.

Let us consider the peak frequency. Assuming the parameter dependence of
Eq.~(\ref{eq5-1}) and replacing the numerical coefficient of ${\cal
O}(1)$ by that of the numerical result, we obtain
\begin{eqnarray}
   f_{\mathrm{peak}} &\simeq& 9\times 10^{10}\times\lambda^{1/2}
    \left(\frac{100}{g_*}\right)^{1/3}
    \left(\frac{t_*}{151t_i}\right)^{1/2}\mathrm{Hz} 
    \nonumber\\
    &\simeq& 9\times 10^9\times \epsilon^{-1/2}\lambda^{3/4}
     \left(\frac{100}{g_*}\right)^{1/12}
     \left(\frac{\eta}{10^{17}\mathrm{GeV}}\right)^{1/2}
     \mathrm{Hz}.
     \label{eq5-5}
\end{eqnarray}
In the second line, we substitute $t_{\mathrm{dec}}$ given by
Eq.~(\ref{eq2-11}) into $t_*$. The difference in $g_*$ dependence
between the first line and the second line comes from 
$t_i = (45M_P^2/16\pi^2g_*T_i^4)^{1/2}\propto g_*^{-1/2}\eta^{-2}$ which is
determined by the Friedmann equation at the initial time. From
Eq.~(\ref{eq5-5}), we expect that the location of the peak shifts to
lower frequency for a larger value of $\epsilon$. This is because if
$\epsilon$ is large, domain walls collapse in the early stage of their
evolution, and the gravitational waves redshifts for a longer time than the
case of a small $\epsilon$. In fact, the peak in our numerical
simulations shifts to the lower frequency for a larger value of
$\epsilon$ (see Fig.~\ref{fig3}). But if $\epsilon$ is sufficiently
large, the peak mixes with the primary spectrum produced at 
$t\lesssim 30t_i$, therefore it is difficult to check precisely the 
$\epsilon$ dependence described by Eq.~(\ref{eq5-5}).

\begin{figure}[htbp]
\begin{center}
\includegraphics[width=60mm,angle=-90]{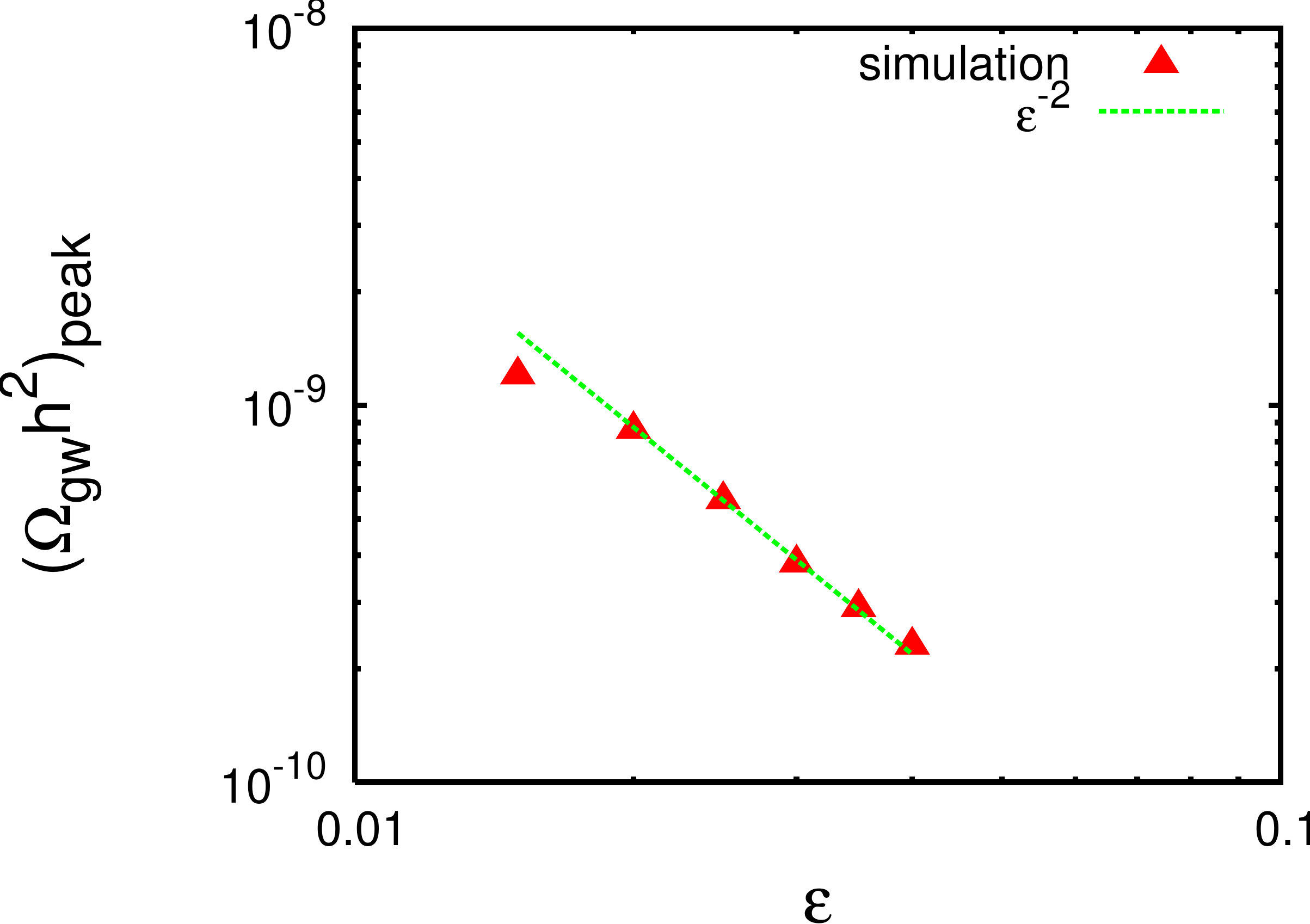}
\end{center}
\caption{The relation between
$(\Omega_{\mathrm{gw}}h^2)_{\mathrm{peak}}$ and $\epsilon$ obtained from
the numerical results. The dotted line represents the fitting formula
given in Eq. (\ref{eq5-6}).}  
\label{fig5}
\end{figure}

Next we consider the peak amplitude of the gravitational
waves. Substituting $t_*\simeq t_{\mathrm{dec}}$ into Eq. (\ref{eq5-2}),
we obtain the parameter dependence of the amplitude
$\Omega_{\mathrm{gw}}h^2\propto
g_*^{-4/3}\lambda\eta^2(t_{\mathrm{dec}}/t_i)^2\propto
g_*^{-1/3}\lambda^2\eta^4\epsilon^{-2}$. So the amplitude is
proportional to $\epsilon^{-2}$. In Fig. \ref{fig5}, we plot the
$\epsilon$ dependence of the peak amplitudes obtained by numerical
results for $b=50$, and we found that the property
$\Omega_{\mathrm{gw}}h^2\propto \epsilon^{-2}$ is indeed observed in our
numerical results. By fitting the numerical coefficients into the line
in Fig.~\ref{fig5}, we obtain the peak amplitude
\begin{equation}
   (\Omega_{\mathrm{gw}}h^2)_{\mathrm{peak}}\simeq 
    3\times 10^{-13}\times\epsilon^{-2}\lambda^{-2}
    \left(\frac{100}{g_*}\right)^{1/3}
    \left(\frac{\eta}{10^{17}\mathrm{GeV}}\right)^4. 
    \label{eq5-6}
\end{equation}

Note that the result with $\epsilon=0.015$ slightly deviates from the
line of $\epsilon^{-2}$ in Fig.~\ref{fig5}. There are some possibilities
to explain this deviation. One explanation is that the deviation is due
to the lack of the simulation time. If $\epsilon$ is sufficiently small,
the domain walls have not completely collapsed in the simulation
time. Therefore, there would be growth of $\Omega_{\mathrm{gw}}h^2$ if
we run the simulation for a longer time. Another possibility is that the
relation $\Omega_{\mathrm{gw}}h^2\propto \epsilon^{-2}$ is not exactly
satisfied for small $\epsilon$. In fact, if we take the limit
$\epsilon\to0$, $\Omega_{\mathrm{gw}}h^2$ diverges. Moreover, we must be
careful about the contribution of the energy density of domain walls to
the total energy density of the universe for the case with small
$\epsilon$, which is ignored here. Unfortunately, we can not test these
possibilities in our current capability of computers.

\begin{figure}[htbp]
\begin{center}
\includegraphics[width=90mm]{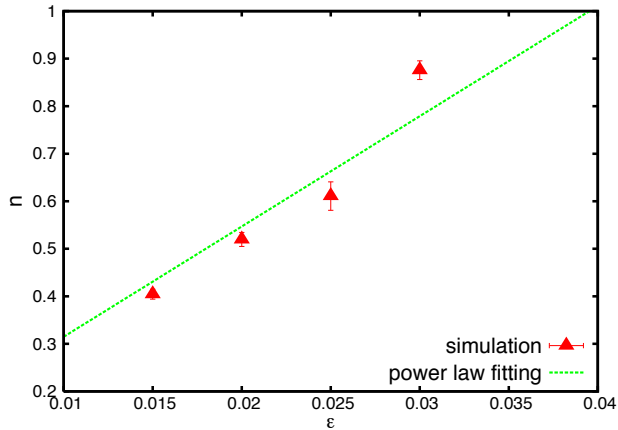}
\end{center}
\caption{The relation between the spectral index $n$ and the bias
$\epsilon$. The dotted line represents the fitting formula given in
Eq.~(\ref{eq5-7}).} 
\label{fig6}
\end{figure}

Now, we consider the intermediate frequency range between the high
frequency peak and the low frequency edge. We parameterize the form of
the spectrum in this range by $\Omega_{\mathrm{gw}}h^2\propto f^n$. In
Fig.~\ref{fig6}, we plot the value of $n$ from numerical results in the
case with the box size $b=50$ for $\epsilon=$0.015, 0.02, 0.025, and
0.03. We also include the data with $\epsilon=0$ for the power law fitting
in Fig.~\ref{fig6}. The result of $\epsilon=0$ is not the spectrum which we 
would observe because the walls do not annihilate in the simulation time. 
However, we expect that the form of the spectrum with $\epsilon \ll 1$ is well 
approximated by the result with $\epsilon=0$ if we assume that the 
low frequency modes of the gravitational waves are generated by 
the scaling evolution of the wall networks.
Note that we omit the data for $\epsilon=0.035$ and 0.04 since the
walls collapse without following scaling phase sufficiently and the form
of the spectrum significantly different from that of the small value of
$\epsilon$. By fitting the numerical results, we obtain the frequency
dependence as
\begin{equation}
   \Omega_{\mathrm{gw}}h^2 \propto f^{20\epsilon+0.08}\qquad 
    \mathrm{for}\qquad f_{\mathrm{edge}}<f<f_{\mathrm{peak}}. 
    \label{eq5-7}
\end{equation}
For a large value of $\epsilon$, the domain walls immediately collapse
after the formation, and there is little growth of the amplitude in the
lower frequency modes corresponding to the horizon scale, resulting in a
steeper spectrum. On the other hand, for a small value of $\epsilon$,
domain walls are straightened up to horizon scale before they
collapse. In this case, the difference in the value of
$\Omega_{\mathrm{gw}}h^2$ between the peak and the edge is determined by
the duration time of the collapse, which might be independent of
$\epsilon$. Therefore, the value of the spectral index is independent of
bias for $\epsilon\ll 1$ and then $n\simeq 0.08$. We emphasize that there might be 
large uncertainties in the parameterization in Eq.~(\ref{eq5-7}). If we write the frequency 
dependence as $\Omega_{\mathrm{gw}}h^2 \propto f^{\alpha\epsilon+\beta}$, the standard 
errors for the coefficients $\alpha$ and $\beta$ are about 10\% and 50\%, respectively. 
These errors arise from the fact that there are only a few data points because of the lack
of the dynamical range of the simulations. It will cause large uncertainties in the 
magnitude of gravitational waves 
relevant to the observations when we extrapolate the result in Section~\ref{sec5D}.

Before closing this subsection, we consider the edge in low
frequency. We can not read the precise location of the edge from the
numerical results because of the large statistical variance in the low
frequency modes. Therefore, here we just derive the parameter dependence
of the frequency and the amplitude for the low frequency edge
analytically, instead of fitting numerical coefficients into the results
of the numerical simulations as the previous discussions. Substituting
$t=t_*\simeq t_{\mathrm{dec}}$ into Eq. (\ref{eq5-3}), we find
\begin{equation}
   f_{\mathrm{edge}} = f_{\mathrm{hor}}(t_{\mathrm{dec}}) 
    = 5 \times 10^{10}\times \epsilon^{1/2}\lambda^{-1/4}
    \left(\frac{100}{g_*}\right)^{1/12}
    \left(\frac{\eta}{10^{17}\mathrm{GeV}}\right)^{1/2}
    \mathrm{Hz}. 
    \label{eq5-8}
\end{equation}
For small $\epsilon$, the life time of domain walls becomes long, and
the horizon scale at $t_{\mathrm{dec}}$ becomes large. Then
$f_{\mathrm{edge}}$ shifts to the lower frequency, as shown in $b=50$
case of Fig.~\ref{fig3}.

Also, by using Eq.~(\ref{eq5-7}), we can estimate the amplitude of
gravitational waves at the edge as
\begin{equation}
   (\Omega_{\mathrm{gw}}h^2)_{\mathrm{edge}} \simeq 
    \left(\frac{f_{\mathrm{edge}}}{f_{\mathrm{peak}}}
    \right)^{20\epsilon+0.08}(\Omega_{\mathrm{gw}}h^2)_{\mathrm{peak}}. 
    \label{eq5-9}
\end{equation}
Substituting Eqs.~(\ref{eq5-5}), (\ref{eq5-6}), and (\ref{eq5-8}) into
Eq.~(\ref{eq5-9}), we obtain
\begin{equation}
   (\Omega_{\mathrm{gw}}h^2)_{\mathrm{edge}} 
    \simeq 3 \times 10^{-13}\times \epsilon^{-2}\lambda^{-2}
    \left(\frac{5.8\epsilon}{\lambda}\right)^{20\epsilon+0.08}
    \left(\frac{100}{g_*}\right)^{1/3}
    \left(\frac{\eta}{10^{17}\mathrm{GeV}}\right)^4.
    \label{eq5-10}
\end{equation}

\subsection{Forecasts for Observations}
\label{sec5D}

In the previous subsection, we found the fitting formulae given by
Eqs.~(\ref{eq5-5})-(\ref{eq5-8}) and (\ref{eq5-10}) by combining the
results of the numerical calculations and the analytic estimations. Now
we use them to predict the gravitational wave spectrum for generic
values of parameters. For simplicity we take $\lambda=1.0$ and $g_*=100$
in the following, then the spectrum depends on two parameters, the mass
scale of the scalar field $\eta$ and the bias $\epsilon$. Note that
$\epsilon$ must satisfy the condition given by Eq.~(\ref{eq4-3}) so that
our results are applicable.

At a glance of the formulae~(\ref{eq5-5})-(\ref{eq5-10}), we can say
that $\eta$ determines the frequency of the spectrum and the strength of
the amplitude, while $\epsilon$ determines the amplitude and the
bandwidth of the gravitational wave spectrum. The large value of $\eta$
results in the appearance of the gravitational wave spectrum in high frequency
range and large $\Omega_{\mathrm{gw}}h^2$. The small value of $\epsilon$
results in the broad band of spectrum and also large
$\Omega_{\mathrm{gw}}h^2$. In Table \ref{tab1}, we show some examples of
the value of various quantities estimated by using
Eqs.~(\ref{eq5-5})-(\ref{eq5-10}).

\begin{table}[b]
\begin{center} 
\caption{The estimation of the amplitude and the frequency of the peak
and the edge for various value of $\eta$. We also show the condition for
$\epsilon$ given by Eq. (\ref{eq4-3}). The upper line in each row
corresponds to the value in the case where $\epsilon$ is taken as its
minimum value, and the lower line corresponds to the value in the case
where $\epsilon$ is taken as its maximum value. Note that domain walls with 
a low energy scale such as $\eta=100$GeV and small $\epsilon$ might be
ruled out by the pulsar timing experiments (see Fig.~\ref{fig7}).}

\vspace{3mm}
\begin{tabular}{c c c c c c}
\hline\hline
$\eta$ & $f_{\mathrm{peak}}$ &
  $(\Omega_{\mathrm{gw}}h^2)_{\mathrm{peak}}$ 
  & $f_{\mathrm{edge}}$ & $(\Omega_{\mathrm{gw}}h^2)_{\mathrm{edge}}$ 
  & Condition for $\epsilon$\\
\hline\hline
\multirow{2}{*}{$10^{15}$GeV} 
  & $4.5\times10^{12}$Hz  & $2.1\times10^{-6}$ 
  & $9.8\times10^5$Hz  & $6.0\times10^{-7}$ 
  & \multirow{2}{*}{$3.7\times10^{-8}<\epsilon<3.8\times10^{-3}$}\\
  & $1.4\times10^{10}$Hz & $1.9\times10^{-16}$ 
  & $3.1\times10^8$Hz & $1.1\times10^{-16}$ & \\
\hline
\multirow{2}{*}{$10^{12}$GeV} 
  & $1.4\times10^{14}$Hz & $2.1\times10^{-6}$ 
  & $31$Hz  & $2.0\times10^{-7}$ & 
\multirow{2}{*}{$3.7\times10^{-14}<\epsilon<3.8\times10^{-6}$}\\
  & $1.4\times10^{10}$Hz & $1.9\times10^{-22}$ 
  & $3.1\times10^5$Hz & $8.2\times10^{-23}$ & \\
\hline
\multirow{2}{*}{$10^{10}$GeV} 
  & $1.4\times10^{15}$Hz & $2.1\times10^{-6}$ 
  & $3.1\times10^{-2}$Hz  & $9.5\times10^{-8}$ & 
\multirow{2}{*}{$3.7\times10^{-18}<\epsilon<3.8\times10^{-8}$}\\
  & $1.4\times10^{10}$Hz & $1.9\times10^{-26}$ 
  & $3.1\times10^3$Hz & $5.7\times10^{-27}$ & \\
\hline
\multirow{2}{*}{$100$GeV} 
  & $1.4\times10^{19}$Hz & $2.1\times10^{-6}$ 
  & $3.1\times10^{-14}$Hz  & $5.0\times10^{-9}$  
  & \multirow{2}{*}{$3.7\times10^{-34}<\epsilon<3.8\times10^{-16}$}\\
  & $1.4\times10^{10}$Hz & $1.9\times10^{-42}$ &
   $3.1\times10^{-5}$Hz & $1.3\times10^{-43}$ & \\
\hline\hline
\label{tab1}
\end{tabular}
\end{center}
\end{table}

\begin{figure}[htbp]
\begin{center}
\includegraphics[scale=0.55]{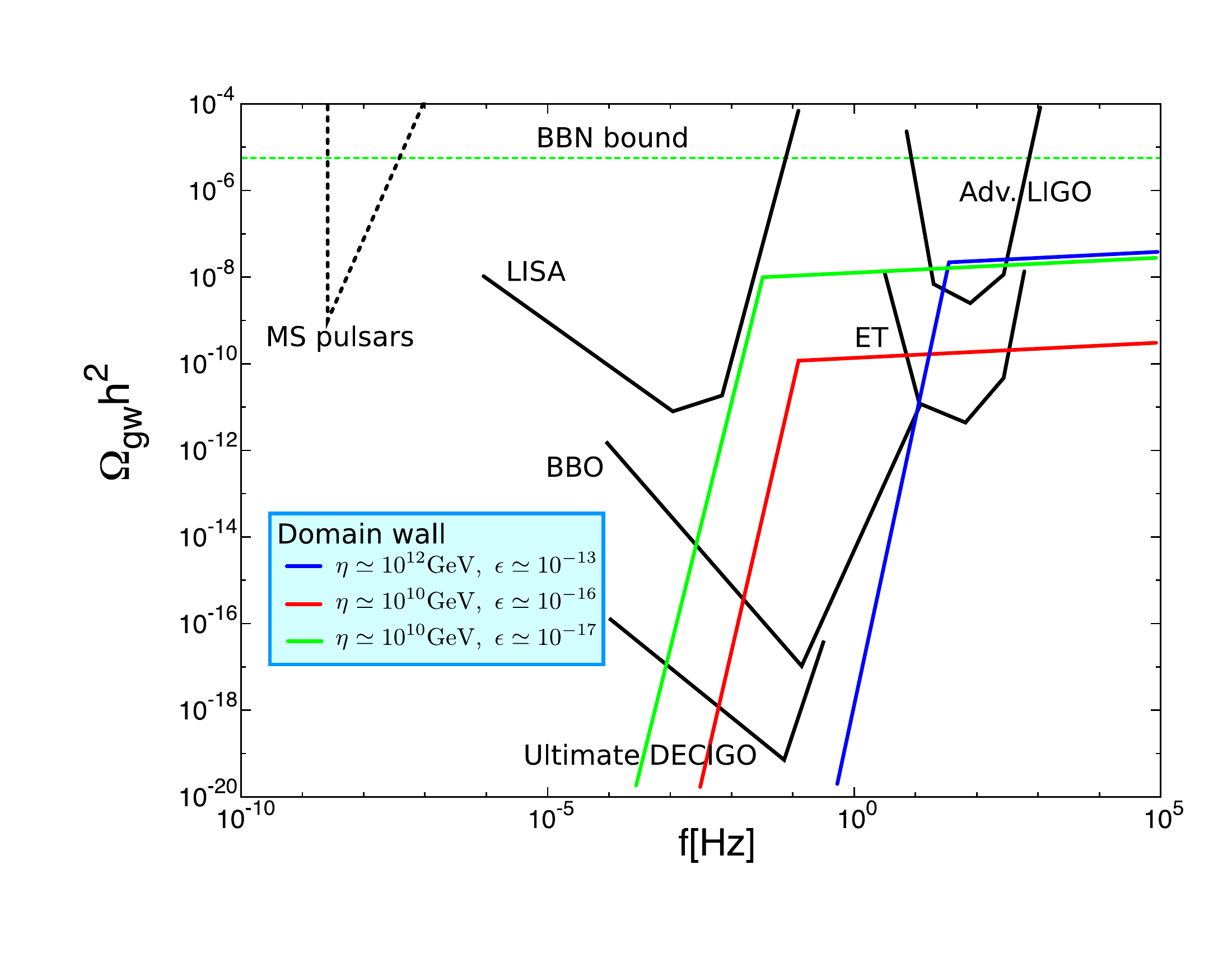}
\end{center}
\caption{The schematics of gravitational wave spectra from collapsing
domain walls. We show three spectra for the case with $(\eta,
\epsilon)=(10^{12}\mathrm{GeV}, 10^{-13})$ (blue), $(\eta,
\epsilon)=(10^{10}\mathrm{GeV}, 10^{-16})$ (red), and $(\eta,
\epsilon)=(10^{10}\mathrm{GeV}, 10^{-17})$ (green). We also roughly plot
the sensitivity of planned detectors, Advanced LIGO~\cite{AdvLIGO},
LISA~\cite{LISA}, ET~\cite{ET}, BBO \cite{2005PhRvD..72h3005C}, and
Ultimate DECIGO~\cite{2001PhRvL..87v1103S}. The dotted lines represent
the constraints obtained from the observation of millisec
pulsars~\cite{1994ApJ...428..713K,2002nsps.conf..114L} (black) and Big
Bang Nucleosynthesis~\cite{2000PhR...331..283M,1997stgr.proc....3A}
(green).}  
\label{fig7}
\end{figure}

\begin{figure}[htbp]
\begin{center}
\includegraphics[scale=0.55]{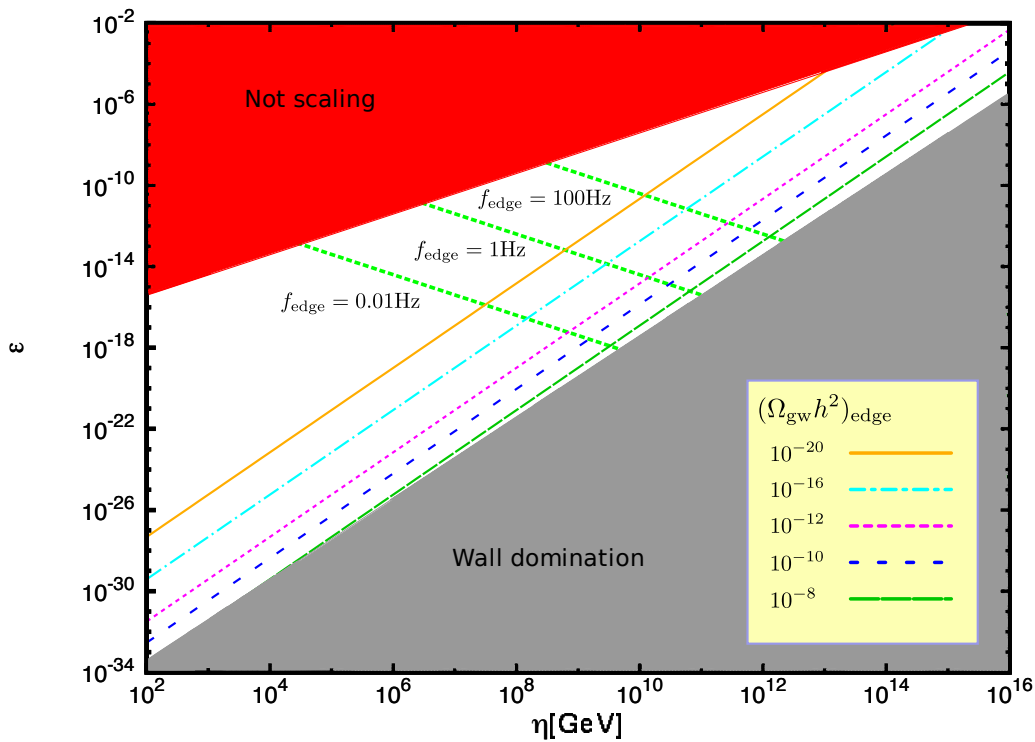}
\end{center}
\caption{The contours of the frequency and the amplitude of the edge in
the parameter space of $\eta$ and $\epsilon$ (we fixed $\lambda=1.0$ and
$g_*=100$). The gray region corresponds to the case that the energy
density of domain walls dominates the total energy density of the
universe [the left hand side of Eq.~(\ref{eq4-3})]. The red region
corresponds to the case that $\epsilon$ is too large to form domain
walls with scaling phase [the right hand side of Eq.~(\ref{eq4-3})]. The
dotted green line shows the frequency $f_{\mathrm{edge}}=0.01$Hz, 1Hz,
and 100Hz. We also show the line with the amplitudes of the edge,
$\Omega_{\mathrm{gw}}h^2=10^{-20}$ (orange), $10^{-16}$ (light blue),
$10^{-12}$ (pink), $10^{-10}$ (blue), and $10^{-8}$ (light green).}
\label{fig8}
\end{figure}

From Table \ref{tab1}, we see that if the value of $\eta$ is around
$10^{10}$-$10^{12}$GeV, the edge would appear in the frequency band
observable in future gravitational wave astronomy projects such as ET, BBO,
and DECIGO. On the other hand, the peak of spectra appears in too high
frequency to be observable. In Fig.~\ref{fig7}, we show examples of
gravitational wave spectra from domain walls estimated by using the
formulae (\ref{eq5-5})-(\ref{eq5-10}). We note that the maximum value of
$\Omega_{\mathrm{gw}}h^2$ at the peak is smaller than the bound which
comes from Big Bang Nucleosynthesis
(BBN)~\cite{2000PhR...331..283M,1997stgr.proc....3A}. We also show the
contours of the frequency and the amplitude of the edge in the parameter
space of $\eta$ and $\epsilon$ in Fig. \ref{fig8}. The parameter
$\epsilon$ is constrained by the condition given by Eq.~(\ref{eq4-3}),
and the white region in Fig.~\ref{fig8} is allowed by this
constraint. The red region denoted by ``Not scaling'' is the parameter
space in which walls disappear before they straighten up to horizon
scale. The gray region denoted by ``Wall domination'' is the parameter
space in which the energy density of domain walls dominates the total
energy density of the universe. Note that the red region and the gray
region {\it do not} imply that parameters in these regions are ruled
out. In fact, domain walls can decay before reaching the scaling regime
if $\epsilon$ is sufficiently large. Also, it might be possible that
domain walls dominate the energy density of the universe if they
disappear before the BBN epoch. However, we can use the
formulae~(\ref{eq5-5})-(\ref{eq5-10}) only in the white region because
in our analysis we assumed that domain walls straighten up to the
horizon scale and disappear before they dominates the energy density of
the universe. Especially, our numerical simulations are performed in the
radiation dominated background, and we do not take into account the
contribution of the energy density of domain walls.  From
Fig.~\ref{fig8}, we can say that if $\eta$ is around
$10^{10}$-$10^{12}$GeV, we would be able to measure the value of the
scale of $\eta$ and the bias $\epsilon$ from the frequency and the
amplitude of the edge of the gravitational wave spectra in future
gravitational wave experiments.

Before going to the conclusion, we comment on the accuracy of the estimation we made. 
There are large uncertainties in the determination of the frequency dependence
in Eq.~(\ref{eq5-7}) because of the limited dynamical range of the simulations.
As we noted before, the uncertainty of the spectral index $n$ is about 50\%.
This uncertainty might become larger when we extrapolate the result for other values
of $\epsilon$. Even if we neglect this increase in the uncertainties and only take 
the uncertainty in the numerical result into account, we expect that the
uncertainty in the magnitude of $(\Omega_{\mathrm{gw}}h^2)_{\mathrm{edge}}$ 
for the forecast in Fig.~\ref{fig7} can be a factor of ${\cal O}(10^{\pm1})$.
In order to improve the accuracy of the estimation, we have to run simulations with a much larger 
dynamical range and with many realizations. 

\section{Conclusion}
\label{sec6}

In this paper, we have studied gravitational waves from domain walls with
approximate $Z_2$ symmetry. We have performed three dimensional lattice
simulations and followed the process of creation, evolution and collapse
of domain walls. We have confirmed that stable domain walls evolve to
maintain the scaling solution, and biased domain walls annihilate in the
time scale estimated by Eq.~(\ref{eq2-11}). In numerical simulations we
have solved the field equation directly rather than using the standard
PRS method. It is found that the spectrum of gravitational waves extends
over broad frequency band, contrary to the naive expectation that
gravitational waves have a peak at the frequency corresponding to the
horizon scale as conjectured
in~\cite{1998PhRvL..81.5497G,2008PhLB..664..194T}. The results of
numerical simulations have revealed that the spectrum has the edge
corresponding to the horizon scale at the time when walls collapse and
the peak corresponding to the small scale on the wall typically given by
the wall width. In the intermediate range between edge and peak, the
gravitational wave amplitude increases moderately with frequency,
$\Omega_{\mathrm{gw}}h^2\sim f^{0.08}$.

By combining analytic estimations and numerical results, we have
obtained the formulae~(\ref{eq5-5})-(\ref{eq5-8}) and~(\ref{eq5-10}) for
the frequency and the amplitude of gravitational wave spectra. We have 
extrapolated the results to more generic parameter space and showed that
the edge of the gravitational wave spectrum from domain walls with
energy scale $\eta=10^{10}$-$10^{12}$GeV is observable in future
gravitational wave experiments such as ET, BBO, and DECIGO.
Therefore, we expect that the gravitational waves from the collapse of domain walls 
can be a new observational window to probe the models of high energy physics which have (approximate) discrete symmetry.
This observation may enable us to search the unknown physics through the estimation of the parameters such as 
bias and provide rich information about the theory beyond the standard model of particle physics.

\begin{acknowledgments}
We would like to thank Kazunori Nakayama and Fuminobu Takahashi for
useful discussions. This work is supported by Grant-in-Aid for
Scientific research from the Ministry of Education, Science, Sports, and
Culture (MEXT), Japan, No.14102004 and No.21111006 (M.K.)  and also by
World Premier International Research Center Initiative (WPI Initiative),
MEXT, Japan.
\end{acknowledgments}


\appendix

\section{Numerical Calculations}
\label{secA}

In this appendix we describe the setup for our numerical computations.

\subsection{Formulation}
\label{secA1}

We adopt the formulation similar to that of Yamaguchi et
al.~\cite{1998PThPh.100..535Y,2003PhRvD..67j3514Y} for the simulation of
global strings. The time, the Hubble radius, and the temperature are
related by the Friedmann equation in the radiation dominated universe
\begin{equation}
   t= \frac{1}{2H} = \frac{\xi}{T^2}, 
    \label{eqA-1}
\end{equation}
where $\xi$ is the constant defined by $\xi\equiv
(45M_P^2/16\pi^2g_*)^{1/2}\simeq 3.68\times
10^{17}(g_*/100)^{-1/2}$GeV. We choose the initial time $t_i$ so that
the temperature at $t=t_i$ is twice of the critical temperature of phase
transition, $T_i=2T_c=4\eta$. We normalize the dimensionful quantities
in the unit of $t_i$. For example, $t\to t/t_i$, $x\to x/t_i$,
$\phi\to\phi t_i$, etc. It is useful to define the dimensionless
quantity $\zeta\equiv \xi/\eta$. By using $\zeta$, equation of motion
for the scalar field renormalized by $t_i$ can be written as
\begin{eqnarray}
    &\displaystyle\ddot{\phi}+\frac{3}{2t}\dot{\phi}
     -\frac{1}{t}\nabla^2\phi + \frac{dV}{d\phi}=0,& 
     \label{eqA-2}\\
    &\displaystyle\frac{dV}{d\phi} = 
     \left(\lambda\phi+\epsilon\frac{\zeta}{16}\right)
     \left(\phi^2-\frac{\zeta^2}{256}\right)
     +\frac{\lambda\zeta^2}{64t}\phi.& 
     \label{eqA-3}
\end{eqnarray}
We choose the value of $\zeta$ as 3.5 which corresponds to
$\eta=1.05\times 10^{17}$GeV.

The (comoving) size of the simulation box is chosen to be $b$ in the
unit of $t_i$, and the lattice spacing is $\delta x =b/N$ where $N$ is
the number of grid points. Here we take $N=256$. The ratio between the
Hubble horizon scale and the physical lattice spacing is
\begin{equation}
    \frac{H^{-1}}{\delta x_{\mathrm{phys}}} 
     = \frac{2N}{b}(t/t_i)^{1/2}, 
     \label{eqA-4}
\end{equation}
and the ratio between the wall width and the physical lattice spacing is
\begin{equation}
    \frac{\delta_w}{\delta x_{\mathrm{phys}}} 
     = \frac{16N}{\lambda^{1/2}\zeta b}(t/t_i)^{-1/2}. 
     \label{eqA-5}
\end{equation}
At the final time $t=151t_i$, these ratios become $H^{-1}/\delta
x_{\mathrm{phys}} \simeq 252<N$, $\delta_w/\delta x_{\mathrm{phys}}
\simeq 3.81$ for $b=25$, and $H^{-1}/\delta x_{\mathrm{phys}} \simeq
126<N$, $\delta_w/\delta x_{\mathrm{phys}} \simeq 1.90$ for
$b=50$. Therefore, the resolution of the simulations is sufficiently
good even at the final time.

We put the periodic boundary condition in the configuration of the
scalar field. We solve the time evolution by using the fourth order
Runge-Kutta method.

\subsection{Initial Conditions}
\label{secA2}

As the initial conditions, we assume that the quantum field $\phi$ is in
thermal equilibrium with temperature $T_i\equiv 1/\beta$. The field and
its derivative satisfy the equal-time correlation relations
\begin{eqnarray}
   \langle\beta|\phi(x)\phi(y)|\beta\rangle_{\mathrm{equal-time}} 
    &=& \int\frac{d^3k}{(2\pi)^3}\frac{1}{2E_k}
    [1+2n_k]e^{i{\bf k\cdot(x-y)}}, 
    \label{eqA-6}\\
    \langle\beta|\dot{\phi}(x)\dot{\phi}(y)|\beta\rangle_{\mathrm{equal-time}}
    &=& \int \frac{d^3k}{(2\pi)^3}\frac{E_k}{2}
    [1+2n_k]e^{i{\bf k\cdot(x-y)}}, \label{eqA-7}
\end{eqnarray}
where $E_k=\sqrt{k^2+m^2}$ and $n_k$ is the occupation number of the
Bose-Einstein distributions, $n_k=(e^{E_k/T_i}-1)^{-1}$. The mass of the
scalar field is given by
\begin{equation}
   m^2 = \left.\frac{d^2V}{d\phi^2}\right|_{\phi=0} 
    = \frac{\lambda}{4}(T_i^2-4\eta^2) = 3\lambda\eta^2. 
    \label{eqA-8}
\end{equation}
The first term in the square bracket of Eqs.~(\ref{eqA-6}) and
(\ref{eqA-7}) corresponds to the vacuum fluctuations which contribute as
a divergent term when we perform the integral of $k$. Hence we subtract
this term and use the renormalized correlation functions
\begin{eqnarray}
   \langle\beta|\phi(x)\phi(y)|\beta\rangle_{\mathrm{ren.}} 
    &=& \int\frac{d^3k}{(2\pi)^3}\frac{n_k}{E_k}, 
    \label{eqA-9}\\
   \langle\beta|\dot{\phi}(x)\dot{\phi}(y)|\beta\rangle_{\mathrm{ren.}}
    &=& \int\frac{d^3k}{(2\pi)^3}E_kn_k. 
    \label{eqA-10}
\end{eqnarray}
In the momentum space, these correlation functions can be written as
\begin{eqnarray}
   \langle\beta|\tilde{\phi}({\bf k})\tilde{\phi}({\bf k'})
    |\beta\rangle_{\mathrm{ren.}} 
    &=& \frac{n_k}{E_k}(2\pi)^3\delta^{(3)}({\bf k+k'}), 
    \label{eqA-11}\\
   \langle\beta|\dot{\tilde{\phi}}({\bf k})\dot{\tilde{\phi}}({\bf k'})
    |\beta\rangle_{\mathrm{ren.}} 
    &=& E_kn_k(2\pi)^3\delta^{(3)}({\bf k+k'}), 
    \label{eqA-12}
\end{eqnarray}
where $\tilde{\phi}({\bf k})$ is the Fourier transform of $\phi({\bf
x})$. Since $\tilde{\phi}({\bf k})$ and $\dot{\tilde{\phi}}({\bf k})$
are uncorrelated in the momentum space, we generate $\tilde{\phi}({\bf
k})$ and $\dot{\tilde{\phi}}({\bf k})$ in the momentum space randomly
following the Gaussian distribution with
\begin{eqnarray}
    &\displaystyle\langle|\tilde{\phi}({\bf k})|^2\rangle 
    = \frac{n_k}{E_k}V,\qquad 
    \langle|\dot{\tilde{\phi}}({\bf k})|^2\rangle 
    = n_kE_kV,& 
    \label{eqA-13}\\
    &\displaystyle\langle\tilde{\phi}({\bf k})\rangle 
    = \langle\dot{\tilde{\phi}}({\bf k})\rangle = 0.& 
    \label{eqA-14}
\end{eqnarray}
Then we transform them into the configuration space and obtain the initial field configurations
$\phi({\bf x})$ and $\dot{\phi}({\bf x})$. Here we used
$(2\pi)^3\delta^{(3)}(0) \simeq V$, where $V=b^3t_i^3$ is the comoving
volume of the simulation box.

\subsection{Calculation of the Area Density}
\label{secA3}

For calculation of the area density of domain walls, we use the
algorithm introduced by Press, Ryden, and
Spergel~\cite{1989ApJ...347..590P}. Let us call the neighboring grid
points that differ by one in $x$, $y$, $z$ location as the
``link''. Define the quantity $\delta_{\pm}$ which takes the value 1 if
$\phi$ has different signs at the two ends of a link and the value 0 if
$\phi$ has the same sign at the two ends of a link. We evaluate the area
density $A/V$ by summing up $\delta_{\pm}$ for each of the links with
the weighting factor
\begin{equation}
   A/V= C \sum_{\mathrm{links}}\delta_{\pm}
    \frac{|\nabla\phi|}{|\phi_{,x}|+|\phi_{,y}|+|\phi_{,z}|}, 
    \label{eqA-15}
\end{equation}
where $\phi_{,x}$, etc. is a derivative of $\phi({\bf x})$ with respect
to $x$, and $C$ is a normalization coefficient. We choose $C$ such that
$A/V=1$ is satisfied when all the links have the value $\delta_{\pm}=1$.



\begin{thebibliography}{50}
\expandafter\ifx\csname natexlab\endcsname\relax\def\natexlab#1{#1}\fi
\expandafter\ifx\csname bibnamefont\endcsname\relax
  \def\bibnamefont#1{#1}\fi
\expandafter\ifx\csname bibfnamefont\endcsname\relax
  \def\bibfnamefont#1{#1}\fi
\expandafter\ifx\csname citenamefont\endcsname\relax
  \def\citenamefont#1{#1}\fi
\expandafter\ifx\csname url\endcsname\relax
  \def\url#1{\texttt{#1}}\fi
\expandafter\ifx\csname urlprefix\endcsname\relax\def\urlprefix{URL }\fi
\providecommand{\bibinfo}[2]{#2}
\providecommand{\eprint}[2][]{\url{#2}}

\bibitem[{\citenamefont{{Kibble}}(1976)}]{1976JPhA....9.1387K}
\bibinfo{author}{\bibfnamefont{T.~W.~B.} \bibnamefont{{Kibble}}},
  \bibinfo{journal}{Journal of Physics A Mathematical General}
  \textbf{\bibinfo{volume}{9}}, \bibinfo{pages}{1387} (\bibinfo{year}{1976}).

\bibitem[{\citenamefont{{Vilenkin} and {Shellard}}(1994)}]{1994csot.book.....V}
\bibinfo{author}{\bibfnamefont{A.}~\bibnamefont{{Vilenkin}}} \bibnamefont{and}
  \bibinfo{author}{\bibfnamefont{E.~P.~S.} \bibnamefont{{Shellard}}},
  \emph{\bibinfo{title}{{Cosmic strings and other topological defects}}}
  (\bibinfo{publisher}{Cambridge University Press}, \bibinfo{year}{1994}).

\bibitem[{\citenamefont{{Gelmini} et~al.}(1989)\citenamefont{{Gelmini},
  {Gleiser}, and {Kolb}}}]{1989PhRvD..39.1558G}
\bibinfo{author}{\bibfnamefont{G.~B.} \bibnamefont{{Gelmini}}},
  \bibinfo{author}{\bibfnamefont{M.}~\bibnamefont{{Gleiser}}},
  \bibnamefont{and} \bibinfo{author}{\bibfnamefont{E.~W.}
  \bibnamefont{{Kolb}}}, \bibinfo{journal}{\prd} \textbf{\bibinfo{volume}{39}},
  \bibinfo{pages}{1558} (\bibinfo{year}{1989}).

\bibitem[{\citenamefont{{Zel'Dovich} et~al.}(1974)\citenamefont{{Zel'Dovich},
  {Kobzarev}, and {Okun'}}}]{1974JETP...40....1Z}
\bibinfo{author}{\bibfnamefont{Y.~B.} \bibnamefont{{Zel'Dovich}}},
  \bibinfo{author}{\bibfnamefont{I.~Y.} \bibnamefont{{Kobzarev}}},
  \bibnamefont{and} \bibinfo{author}{\bibfnamefont{L.~B.}
  \bibnamefont{{Okun'}}}, \bibinfo{journal}{Soviet Journal of Experimental and
  Theoretical Physics} \textbf{\bibinfo{volume}{40}}, \bibinfo{pages}{1}
  (\bibinfo{year}{1974}).

\bibitem[{\citenamefont{{Rai} and
  {Senjanovi{\'c}}}(1994)}]{1994PhRvD..49.2729R}
\bibinfo{author}{\bibfnamefont{B.}~\bibnamefont{{Rai}}} \bibnamefont{and}
  \bibinfo{author}{\bibfnamefont{G.}~\bibnamefont{{Senjanovi{\'c}}}},
  \bibinfo{journal}{\prd} \textbf{\bibinfo{volume}{49}}, \bibinfo{pages}{2729}
  (\bibinfo{year}{1994}), \eprint{arXiv:hep-ph/9301240}.

\bibitem[{\citenamefont{{Vilenkin}}(1981)}]{1981PhRvD..23..852V}
\bibinfo{author}{\bibfnamefont{A.}~\bibnamefont{{Vilenkin}}},
  \bibinfo{journal}{\prd} \textbf{\bibinfo{volume}{23}}, \bibinfo{pages}{852}
  (\bibinfo{year}{1981}).

\bibitem[{\citenamefont{{Sikivie}}(1982)}]{1982PhRvL..48.1156S}
\bibinfo{author}{\bibfnamefont{P.}~\bibnamefont{{Sikivie}}},
  \bibinfo{journal}{Physical Review Letters} \textbf{\bibinfo{volume}{48}},
  \bibinfo{pages}{1156} (\bibinfo{year}{1982}).

\bibitem[{\citenamefont{{Mohanty} and {Stecker}}(1984)}]{1984PhLB..143..351M}
\bibinfo{author}{\bibfnamefont{A.~K.} \bibnamefont{{Mohanty}}}
  \bibnamefont{and} \bibinfo{author}{\bibfnamefont{F.~W.}
  \bibnamefont{{Stecker}}}, \bibinfo{journal}{Physics Letters B}
  \textbf{\bibinfo{volume}{143}}, \bibinfo{pages}{351} (\bibinfo{year}{1984}).

\bibitem[{\citenamefont{{Lalak} et~al.}(1995)\citenamefont{{Lalak}, {Ovrut},
  and {Thomas}}}]{1995PhRvD..51.5456L}
\bibinfo{author}{\bibfnamefont{Z.}~\bibnamefont{{Lalak}}},
  \bibinfo{author}{\bibfnamefont{B.~A.} \bibnamefont{{Ovrut}}},
  \bibnamefont{and} \bibinfo{author}{\bibfnamefont{S.}~\bibnamefont{{Thomas}}},
  \bibinfo{journal}{\prd} \textbf{\bibinfo{volume}{51}}, \bibinfo{pages}{5456}
  (\bibinfo{year}{1995}).

\bibitem[{\citenamefont{{Coulson} et~al.}(1996)\citenamefont{{Coulson},
  {Lalak}, and {Ovrut}}}]{1996PhRvD..53.4237C}
\bibinfo{author}{\bibfnamefont{D.}~\bibnamefont{{Coulson}}},
  \bibinfo{author}{\bibfnamefont{Z.}~\bibnamefont{{Lalak}}}, \bibnamefont{and}
  \bibinfo{author}{\bibfnamefont{B.}~\bibnamefont{{Ovrut}}},
  \bibinfo{journal}{\prd} \textbf{\bibinfo{volume}{53}}, \bibinfo{pages}{4237}
  (\bibinfo{year}{1996}).

\bibitem[{\citenamefont{{Hawking} et~al.}(1982)\citenamefont{{Hawking}, {Moss},
  and {Stewart}}}]{1982PhRvD..26.2681H}
\bibinfo{author}{\bibfnamefont{S.~W.} \bibnamefont{{Hawking}}},
  \bibinfo{author}{\bibfnamefont{I.~G.} \bibnamefont{{Moss}}},
  \bibnamefont{and} \bibinfo{author}{\bibfnamefont{J.~M.}
  \bibnamefont{{Stewart}}}, \bibinfo{journal}{\prd}
  \textbf{\bibinfo{volume}{26}}, \bibinfo{pages}{2681} (\bibinfo{year}{1982}).

\bibitem[{\citenamefont{{Turner} and {Wilczek}}(1990)}]{1990PhRvL..65.3080T}
\bibinfo{author}{\bibfnamefont{M.~S.} \bibnamefont{{Turner}}} \bibnamefont{and}
  \bibinfo{author}{\bibfnamefont{F.}~\bibnamefont{{Wilczek}}},
  \bibinfo{journal}{Physical Review Letters} \textbf{\bibinfo{volume}{65}},
  \bibinfo{pages}{3080} (\bibinfo{year}{1990}).

\bibitem[{\citenamefont{{Kosowsky}
  et~al.}(1992{\natexlab{a}})\citenamefont{{Kosowsky}, {Turner}, and
  {Watkins}}}]{1992PhRvD..45.4514K}
\bibinfo{author}{\bibfnamefont{A.}~\bibnamefont{{Kosowsky}}},
  \bibinfo{author}{\bibfnamefont{M.~S.} \bibnamefont{{Turner}}},
  \bibnamefont{and}
  \bibinfo{author}{\bibfnamefont{R.}~\bibnamefont{{Watkins}}},
  \bibinfo{journal}{\prd} \textbf{\bibinfo{volume}{45}}, \bibinfo{pages}{4514}
  (\bibinfo{year}{1992}{\natexlab{a}}).

\bibitem[{\citenamefont{{Kosowsky}
  et~al.}(1992{\natexlab{b}})\citenamefont{{Kosowsky}, {Turner}, and
  {Watkins}}}]{1992PhRvL..69.2026K}
\bibinfo{author}{\bibfnamefont{A.}~\bibnamefont{{Kosowsky}}},
  \bibinfo{author}{\bibfnamefont{M.~S.} \bibnamefont{{Turner}}},
  \bibnamefont{and}
  \bibinfo{author}{\bibfnamefont{R.}~\bibnamefont{{Watkins}}},
  \bibinfo{journal}{Physical Review Letters} \textbf{\bibinfo{volume}{69}},
  \bibinfo{pages}{2026} (\bibinfo{year}{1992}{\natexlab{b}}).

\bibitem[{\citenamefont{{Kosowsky} and {Turner}}(1993)}]{1993PhRvD..47.4372K}
\bibinfo{author}{\bibfnamefont{A.}~\bibnamefont{{Kosowsky}}} \bibnamefont{and}
  \bibinfo{author}{\bibfnamefont{M.~S.} \bibnamefont{{Turner}}},
  \bibinfo{journal}{\prd} \textbf{\bibinfo{volume}{47}}, \bibinfo{pages}{4372}
  (\bibinfo{year}{1993}), \eprint{arXiv:astro-ph/9211004}.

\bibitem[{\citenamefont{{Kamionkowski}
  et~al.}(1994)\citenamefont{{Kamionkowski}, {Kosowsky}, and
  {Turner}}}]{1994PhRvD..49.2837K}
\bibinfo{author}{\bibfnamefont{M.}~\bibnamefont{{Kamionkowski}}},
  \bibinfo{author}{\bibfnamefont{A.}~\bibnamefont{{Kosowsky}}},
  \bibnamefont{and} \bibinfo{author}{\bibfnamefont{M.~S.}
  \bibnamefont{{Turner}}}, \bibinfo{journal}{\prd}
  \textbf{\bibinfo{volume}{49}}, \bibinfo{pages}{2837} (\bibinfo{year}{1994}),
  \eprint{arXiv:astro-ph/9310044}.

\bibitem[{\citenamefont{{Gleiser} and {Roberts}}(1998)}]{1998PhRvL..81.5497G}
\bibinfo{author}{\bibfnamefont{M.}~\bibnamefont{{Gleiser}}} \bibnamefont{and}
  \bibinfo{author}{\bibfnamefont{R.}~\bibnamefont{{Roberts}}},
  \bibinfo{journal}{Physical Review Letters} \textbf{\bibinfo{volume}{81}},
  \bibinfo{pages}{5497} (\bibinfo{year}{1998}),
  \eprint{arXiv:astro-ph/9807260}.

\bibitem[{\citenamefont{{Maggiore}}(2000)}]{2000PhR...331..283M}
\bibinfo{author}{\bibfnamefont{M.}~\bibnamefont{{Maggiore}}},
  \bibinfo{journal}{\physrep} \textbf{\bibinfo{volume}{331}},
  \bibinfo{pages}{283} (\bibinfo{year}{2000}), \eprint{arXiv:gr-qc/9909001}.

\bibitem[{\citenamefont{{Maggiore}}(2008)}]{Maggiore2008}
\bibinfo{author}{\bibfnamefont{M.}~\bibnamefont{{Maggiore}}},
  \emph{\bibinfo{title}{{{\it Gravitational Waves Volume 1: Theory and
  Experiments}}}} (\bibinfo{publisher}{Oxford University Press},
  \bibinfo{year}{2008}).

\bibitem[{\citenamefont{{Sathyaprakash} and
  {Schutz}}(2009)}]{2009LRR....12....2S}
\bibinfo{author}{\bibfnamefont{B.~S.} \bibnamefont{{Sathyaprakash}}}
  \bibnamefont{and} \bibinfo{author}{\bibfnamefont{B.~F.}
  \bibnamefont{{Schutz}}}, \bibinfo{journal}{Living Reviews in Relativity}
  \textbf{\bibinfo{volume}{12}}, \bibinfo{pages}{2} (\bibinfo{year}{2009}),
  \eprint{0903.0338}.

\bibitem[{\citenamefont{{Abramovici} et~al.}(1992)\citenamefont{{Abramovici},
  {Althouse}, {Drever}, {Gursel}, {Kawamura}, {Raab}, {Shoemaker}, {Sievers},
  {Spero}, and {Thorne}}}]{1992Sci...256..325A}
\bibinfo{author}{\bibfnamefont{A.}~\bibnamefont{{Abramovici}}},
  \bibinfo{author}{\bibfnamefont{W.~E.} \bibnamefont{{Althouse}}},
  \bibinfo{author}{\bibfnamefont{R.~W.~P.} \bibnamefont{{Drever}}},
  \bibinfo{author}{\bibfnamefont{Y.}~\bibnamefont{{Gursel}}},
  \bibinfo{author}{\bibfnamefont{S.}~\bibnamefont{{Kawamura}}},
  \bibinfo{author}{\bibfnamefont{F.~J.} \bibnamefont{{Raab}}},
  \bibinfo{author}{\bibfnamefont{D.}~\bibnamefont{{Shoemaker}}},
  \bibinfo{author}{\bibfnamefont{L.}~\bibnamefont{{Sievers}}},
  \bibinfo{author}{\bibfnamefont{R.~E.} \bibnamefont{{Spero}}},
  \bibnamefont{and} \bibinfo{author}{\bibfnamefont{K.~S.}
  \bibnamefont{{Thorne}}}, \bibinfo{journal}{Science}
  \textbf{\bibinfo{volume}{256}}, \bibinfo{pages}{325} (\bibinfo{year}{1992}).

\bibitem[{\citenamefont{{Kuroda} et~al.}(2002)\citenamefont{{Kuroda}, {Ohashi},
  {Miyoki}, {Ishizuka}, {Taylor}, {Yamamoto}, {Miyakawa}, {Fujimoto},
  {Kawamura}, {Takahashi} et~al.}}]{2002CQGra..19.1237K}
\bibinfo{author}{\bibfnamefont{K.}~\bibnamefont{{Kuroda}}},
  \bibinfo{author}{\bibfnamefont{M.}~\bibnamefont{{Ohashi}}},
  \bibinfo{author}{\bibfnamefont{S.}~\bibnamefont{{Miyoki}}},
  \bibinfo{author}{\bibfnamefont{H.}~\bibnamefont{{Ishizuka}}},
  \bibinfo{author}{\bibfnamefont{C.~T.} \bibnamefont{{Taylor}}},
  \bibinfo{author}{\bibfnamefont{K.}~\bibnamefont{{Yamamoto}}},
  \bibinfo{author}{\bibfnamefont{O.}~\bibnamefont{{Miyakawa}}},
  \bibinfo{author}{\bibfnamefont{M.}~\bibnamefont{{Fujimoto}}},
  \bibinfo{author}{\bibfnamefont{S.}~\bibnamefont{{Kawamura}}},
  \bibinfo{author}{\bibfnamefont{R.}~\bibnamefont{{Takahashi}}},
  \bibnamefont{et~al.}, \bibinfo{journal}{Classical and Quantum Gravity}
  \textbf{\bibinfo{volume}{19}}, \bibinfo{pages}{1237} (\bibinfo{year}{2002}).

\bibitem[{LIS()}]{LISA}
\urlprefix\url{http://lisa.nasa.gov/}.

\bibitem[{\citenamefont{{Crowder} and {Cornish}}(2005)}]{2005PhRvD..72h3005C}
\bibinfo{author}{\bibfnamefont{J.}~\bibnamefont{{Crowder}}} \bibnamefont{and}
  \bibinfo{author}{\bibfnamefont{N.~J.} \bibnamefont{{Cornish}}},
  \bibinfo{journal}{\prd} \textbf{\bibinfo{volume}{72}},
  \bibinfo{pages}{083005} (\bibinfo{year}{2005}), \eprint{arXiv:gr-qc/0506015}.

\bibitem[{\citenamefont{{Kawamura} et~al.}(2006)\citenamefont{{Kawamura},
  {Nakamura}, {Ando}, {Seto}, {Tsubono}, {Numata}, {Takahashi}, {Nagano},
  {Ishikawa}, {Musha} et~al.}}]{2006CQGra..23S.125K}
\bibinfo{author}{\bibfnamefont{S.}~\bibnamefont{{Kawamura}}},
  \bibinfo{author}{\bibfnamefont{T.}~\bibnamefont{{Nakamura}}},
  \bibinfo{author}{\bibfnamefont{M.}~\bibnamefont{{Ando}}},
  \bibinfo{author}{\bibfnamefont{N.}~\bibnamefont{{Seto}}},
  \bibinfo{author}{\bibfnamefont{K.}~\bibnamefont{{Tsubono}}},
  \bibinfo{author}{\bibfnamefont{K.}~\bibnamefont{{Numata}}},
  \bibinfo{author}{\bibfnamefont{R.}~\bibnamefont{{Takahashi}}},
  \bibinfo{author}{\bibfnamefont{S.}~\bibnamefont{{Nagano}}},
  \bibinfo{author}{\bibfnamefont{T.}~\bibnamefont{{Ishikawa}}},
  \bibinfo{author}{\bibfnamefont{M.}~\bibnamefont{{Musha}}},
  \bibnamefont{et~al.}, \bibinfo{journal}{Classical and Quantum Gravity}
  \textbf{\bibinfo{volume}{23}}, \bibinfo{pages}{125} (\bibinfo{year}{2006}).

\bibitem[{ET()}]{ET}
\urlprefix\url{http://www.et-gw.eu/}.

\bibitem[{\citenamefont{{Takahashi} et~al.}(2008)\citenamefont{{Takahashi},
  {Yanagida}, and {Yonekura}}}]{2008PhLB..664..194T}
\bibinfo{author}{\bibfnamefont{F.}~\bibnamefont{{Takahashi}}},
  \bibinfo{author}{\bibfnamefont{T.~T.} \bibnamefont{{Yanagida}}},
  \bibnamefont{and}
  \bibinfo{author}{\bibfnamefont{K.}~\bibnamefont{{Yonekura}}},
  \bibinfo{journal}{Physics Letters B} \textbf{\bibinfo{volume}{664}},
  \bibinfo{pages}{194} (\bibinfo{year}{2008}), \eprint{0802.4335}.

\bibitem[{\citenamefont{{Press} et~al.}(1989)\citenamefont{{Press}, {Ryden},
  and {Spergel}}}]{1989ApJ...347..590P}
\bibinfo{author}{\bibfnamefont{W.~H.} \bibnamefont{{Press}}},
  \bibinfo{author}{\bibfnamefont{B.~S.} \bibnamefont{{Ryden}}},
  \bibnamefont{and} \bibinfo{author}{\bibfnamefont{D.~N.}
  \bibnamefont{{Spergel}}}, \bibinfo{journal}{\apj}
  \textbf{\bibinfo{volume}{347}}, \bibinfo{pages}{590} (\bibinfo{year}{1989}).

\bibitem[{\citenamefont{{Kawano}}(1990)}]{1990PhRvD..41.1013K}
\bibinfo{author}{\bibfnamefont{L.}~\bibnamefont{{Kawano}}},
  \bibinfo{journal}{\prd} \textbf{\bibinfo{volume}{41}}, \bibinfo{pages}{1013}
  (\bibinfo{year}{1990}).

\bibitem[{\citenamefont{{Nambu}}(1970)}]{1970sqm..conf..269N}
\bibinfo{author}{\bibfnamefont{Y.}~\bibnamefont{{Nambu}}}, in
  \emph{\bibinfo{booktitle}{Symmetries and Quark Models}}, edited by
  \bibinfo{editor}{\bibnamefont{{R.~Chand}}} (\bibinfo{year}{1970}), pp.
  \bibinfo{pages}{269--+}.

\bibitem[{\citenamefont{{Got{\= o}}}(1971)}]{1971PThPh..46.1560G}
\bibinfo{author}{\bibfnamefont{T.}~\bibnamefont{{Got{\= o}}}},
  \bibinfo{journal}{Progress of Theoretical Physics}
  \textbf{\bibinfo{volume}{46}}, \bibinfo{pages}{1560} (\bibinfo{year}{1971}).

\bibitem[{\citenamefont{{Garagounis} and
  {Hindmarsh}}(2003)}]{2003PhRvD..68j3506G}
\bibinfo{author}{\bibfnamefont{T.}~\bibnamefont{{Garagounis}}}
  \bibnamefont{and}
  \bibinfo{author}{\bibfnamefont{M.}~\bibnamefont{{Hindmarsh}}},
  \bibinfo{journal}{\prd} \textbf{\bibinfo{volume}{68}},
  \bibinfo{pages}{103506} (\bibinfo{year}{2003}),
  \eprint{arXiv:hep-ph/0212359}.

\bibitem[{\citenamefont{{Oliveira} et~al.}(2005)\citenamefont{{Oliveira},
  {Martins}, and {Avelino}}}]{2005PhRvD..71h3509O}
\bibinfo{author}{\bibfnamefont{J.~C.} \bibnamefont{{Oliveira}}},
  \bibinfo{author}{\bibfnamefont{C.~J.} \bibnamefont{{Martins}}},
  \bibnamefont{and} \bibinfo{author}{\bibfnamefont{P.~P.}
  \bibnamefont{{Avelino}}}, \bibinfo{journal}{\prd}
  \textbf{\bibinfo{volume}{71}}, \bibinfo{pages}{083509}
  (\bibinfo{year}{2005}), \eprint{arXiv:hep-ph/0410356}.

\bibitem[{\citenamefont{{Avelino}
  et~al.}(2005{\natexlab{a}})\citenamefont{{Avelino}, {Oliveira}, and
  {Martins}}}]{2005PhLB..610....1A}
\bibinfo{author}{\bibfnamefont{P.~P.} \bibnamefont{{Avelino}}},
  \bibinfo{author}{\bibfnamefont{J.~C.~R.~E.} \bibnamefont{{Oliveira}}},
  \bibnamefont{and} \bibinfo{author}{\bibfnamefont{C.~J.~A.~P.}
  \bibnamefont{{Martins}}}, \bibinfo{journal}{Physics Letters B}
  \textbf{\bibinfo{volume}{610}}, \bibinfo{pages}{1}
  (\bibinfo{year}{2005}{\natexlab{a}}), \eprint{arXiv:hep-th/0503226}.

\bibitem[{\citenamefont{{Larsson} et~al.}(1997)\citenamefont{{Larsson},
  {Sarkar}, and {White}}}]{1997PhRvD..55.5129L}
\bibinfo{author}{\bibfnamefont{S.~E.} \bibnamefont{{Larsson}}},
  \bibinfo{author}{\bibfnamefont{S.}~\bibnamefont{{Sarkar}}}, \bibnamefont{and}
  \bibinfo{author}{\bibfnamefont{P.~L.} \bibnamefont{{White}}},
  \bibinfo{journal}{\prd} \textbf{\bibinfo{volume}{55}}, \bibinfo{pages}{5129}
  (\bibinfo{year}{1997}), \eprint{arXiv:hep-ph/9608319}.

\bibitem[{\citenamefont{{Hindmarsh}}(1996)}]{1996PhRvL..77.4495H}
\bibinfo{author}{\bibfnamefont{M.}~\bibnamefont{{Hindmarsh}}},
  \bibinfo{journal}{Physical Review Letters} \textbf{\bibinfo{volume}{77}},
  \bibinfo{pages}{4495} (\bibinfo{year}{1996}), \eprint{arXiv:hep-ph/9605332}.

\bibitem[{\citenamefont{{Hindmarsh}}(2003)}]{2003PhRvD..68d3510H}
\bibinfo{author}{\bibfnamefont{M.}~\bibnamefont{{Hindmarsh}}},
  \bibinfo{journal}{\prd} \textbf{\bibinfo{volume}{68}},
  \bibinfo{pages}{043510} (\bibinfo{year}{2003}),
  \eprint{arXiv:hep-ph/0207267}.

\bibitem[{\citenamefont{{Avelino}
  et~al.}(2005{\natexlab{b}})\citenamefont{{Avelino}, {Martins}, and
  {Oliveira}}}]{2005PhRvD..72h3506A}
\bibinfo{author}{\bibfnamefont{P.~P.} \bibnamefont{{Avelino}}},
  \bibinfo{author}{\bibfnamefont{C.~J.~A.~P.} \bibnamefont{{Martins}}},
  \bibnamefont{and} \bibinfo{author}{\bibfnamefont{J.~C.~R.~E.}
  \bibnamefont{{Oliveira}}}, \bibinfo{journal}{\prd}
  \textbf{\bibinfo{volume}{72}}, \bibinfo{pages}{083506}
  (\bibinfo{year}{2005}{\natexlab{b}}), \eprint{arXiv:hep-ph/0507272}.

\bibitem[{\citenamefont{{Stauffer}}(1979)}]{1979PhR....54....1S}
\bibinfo{author}{\bibfnamefont{D.}~\bibnamefont{{Stauffer}}},
  \bibinfo{journal}{\physrep} \textbf{\bibinfo{volume}{54}}, \bibinfo{pages}{1}
  (\bibinfo{year}{1979}).

\bibitem[{\citenamefont{{Dufaux} et~al.}(2007)\citenamefont{{Dufaux},
  {Bergman}, {Felder}, {Kofman}, and {Uzan}}}]{2007PhRvD..76l3517D}
\bibinfo{author}{\bibfnamefont{J.}~\bibnamefont{{Dufaux}}},
  \bibinfo{author}{\bibfnamefont{A.}~\bibnamefont{{Bergman}}},
  \bibinfo{author}{\bibfnamefont{G.}~\bibnamefont{{Felder}}},
  \bibinfo{author}{\bibfnamefont{L.}~\bibnamefont{{Kofman}}}, \bibnamefont{and}
  \bibinfo{author}{\bibfnamefont{J.}~\bibnamefont{{Uzan}}},
  \bibinfo{journal}{\prd} \textbf{\bibinfo{volume}{76}},
  \bibinfo{pages}{123517} (\bibinfo{year}{2007}), \eprint{0707.0875}.

\bibitem[{\citenamefont{{Dufaux} et~al.}(2009)\citenamefont{{Dufaux}, {Felder},
  {Kofman}, and {Navros}}}]{2009JCAP...03..001D}
\bibinfo{author}{\bibfnamefont{J.}~\bibnamefont{{Dufaux}}},
  \bibinfo{author}{\bibfnamefont{G.}~\bibnamefont{{Felder}}},
  \bibinfo{author}{\bibfnamefont{L.}~\bibnamefont{{Kofman}}}, \bibnamefont{and}
  \bibinfo{author}{\bibfnamefont{O.}~\bibnamefont{{Navros}}},
  \bibinfo{journal}{Journal of Cosmology and Astro-Particle Physics}
  \textbf{\bibinfo{volume}{3}}, \bibinfo{pages}{1} (\bibinfo{year}{2009}),
  \eprint{0812.2917}.

\bibitem[{\citenamefont{{Kofman} et~al.}(1994)\citenamefont{{Kofman}, {Linde},
  and {Starobinsky}}}]{1994PhRvL..73.3195K}
\bibinfo{author}{\bibfnamefont{L.}~\bibnamefont{{Kofman}}},
  \bibinfo{author}{\bibfnamefont{A.}~\bibnamefont{{Linde}}}, \bibnamefont{and}
  \bibinfo{author}{\bibfnamefont{A.~A.} \bibnamefont{{Starobinsky}}},
  \bibinfo{journal}{Physical Review Letters} \textbf{\bibinfo{volume}{73}},
  \bibinfo{pages}{3195} (\bibinfo{year}{1994}), \eprint{arXiv:hep-th/9405187}.

\bibitem[{\citenamefont{{Kofman} et~al.}(1997)\citenamefont{{Kofman}, {Linde},
  and {Starobinsky}}}]{1997PhRvD..56.3258K}
\bibinfo{author}{\bibfnamefont{L.}~\bibnamefont{{Kofman}}},
  \bibinfo{author}{\bibfnamefont{A.}~\bibnamefont{{Linde}}}, \bibnamefont{and}
  \bibinfo{author}{\bibfnamefont{A.~A.} \bibnamefont{{Starobinsky}}},
  \bibinfo{journal}{\prd} \textbf{\bibinfo{volume}{56}}, \bibinfo{pages}{3258}
  (\bibinfo{year}{1997}), \eprint{arXiv:hep-ph/9704452}.

\bibitem[{Adv()}]{AdvLIGO}
\urlprefix\url{http://www.ligo.caltech.edu/advLIGO/scripts/summary.shtml}.

\bibitem[{\citenamefont{{Seto} et~al.}(2001)\citenamefont{{Seto}, {Kawamura},
  and {Nakamura}}}]{2001PhRvL..87v1103S}
\bibinfo{author}{\bibfnamefont{N.}~\bibnamefont{{Seto}}},
  \bibinfo{author}{\bibfnamefont{S.}~\bibnamefont{{Kawamura}}},
  \bibnamefont{and}
  \bibinfo{author}{\bibfnamefont{T.}~\bibnamefont{{Nakamura}}},
  \bibinfo{journal}{Physical Review Letters} \textbf{\bibinfo{volume}{87}},
  \bibinfo{pages}{221103} (\bibinfo{year}{2001}),
  \eprint{arXiv:astro-ph/0108011}.

\bibitem[{\citenamefont{{Kaspi} et~al.}(1994)\citenamefont{{Kaspi}, {Taylor},
  and {Ryba}}}]{1994ApJ...428..713K}
\bibinfo{author}{\bibfnamefont{V.~M.} \bibnamefont{{Kaspi}}},
  \bibinfo{author}{\bibfnamefont{J.~H.} \bibnamefont{{Taylor}}},
  \bibnamefont{and} \bibinfo{author}{\bibfnamefont{M.~F.}
  \bibnamefont{{Ryba}}}, \bibinfo{journal}{\apj}
  \textbf{\bibinfo{volume}{428}}, \bibinfo{pages}{713} (\bibinfo{year}{1994}).

\bibitem[{\citenamefont{{Lommen}}(2002)}]{2002nsps.conf..114L}
\bibinfo{author}{\bibfnamefont{A.~N.} \bibnamefont{{Lommen}}}, in
  \emph{\bibinfo{booktitle}{Neutron Stars, Pulsars, and Supernova Remnants}},
  edited by \bibinfo{editor}{\bibnamefont{{W.~Becker, H.~Lesch, \&
  J.~Tr{\"u}mper}}} (\bibinfo{year}{2002}), pp. \bibinfo{pages}{114--+}.

\bibitem[{\citenamefont{{Allen}}(1997)}]{1997stgr.proc....3A}
\bibinfo{author}{\bibfnamefont{B.}~\bibnamefont{{Allen}}}, in
  \emph{\bibinfo{booktitle}{Some Topics on General Relativity and Gravitational
  Radiation}}, edited by \bibinfo{editor}{\bibnamefont{{J.~A.~Miralles,
  J.~A.~Morales, \& D.~Saez}}} (\bibinfo{year}{1997}), pp.
  \bibinfo{pages}{3--+}, \eprint{arXiv:gr-qc/9604033}.

\bibitem[{\citenamefont{{Yamaguchi} et~al.}(1998)\citenamefont{{Yamaguchi},
  {Yokoyama}, and {Kawasaki}}}]{1998PThPh.100..535Y}
\bibinfo{author}{\bibfnamefont{M.}~\bibnamefont{{Yamaguchi}}},
  \bibinfo{author}{\bibfnamefont{J.}~\bibnamefont{{Yokoyama}}},
  \bibnamefont{and}
  \bibinfo{author}{\bibfnamefont{M.}~\bibnamefont{{Kawasaki}}},
  \bibinfo{journal}{Progress of Theoretical Physics}
  \textbf{\bibinfo{volume}{100}}, \bibinfo{pages}{535} (\bibinfo{year}{1998}),
  \eprint{arXiv:hep-ph/9808326}.

\bibitem[{\citenamefont{{Yamaguchi} and
  {Yokoyama}}(2003)}]{2003PhRvD..67j3514Y}
\bibinfo{author}{\bibfnamefont{M.}~\bibnamefont{{Yamaguchi}}} \bibnamefont{and}
  \bibinfo{author}{\bibfnamefont{J.}~\bibnamefont{{Yokoyama}}},
  \bibinfo{journal}{\prd} \textbf{\bibinfo{volume}{67}},
  \bibinfo{pages}{103514} (\bibinfo{year}{2003}),
  \eprint{arXiv:hep-ph/0210343}.

\end{thebibliography}

\end{document}